\documentclass[aps,prx,superscriptaddress,twocolumn]{revtex4}
\usepackage{graphicx}
\usepackage{times}
\usepackage{amsmath}
\usepackage{amsthm}
\usepackage{amssymb}
\usepackage{amsbsy}
\usepackage{tabularx}
\newcommand{\abs}[1]{\vert #1 \vert}
\newcommand{\bra}[1]{\langle #1 \vert}
\newcommand{\ket}[1]{\vert #1 \rangle}

\newcommand{\asinh}{\text{asinh}}

\begin{document}
\title{A practical phase gate for producing Bell violations in Majorana wires}
\author{David J. Clarke}
\affiliation{Condensed Matter Theory Center, Department of Physics, University of Maryland, College Park}
\affiliation{Joint Quantum Institute, University of Maryland, College Park, Maryland 20742, USA}
\author{Jay D. Sau}
\affiliation{Condensed Matter Theory Center, Department of Physics, University of Maryland, College Park}
\affiliation{Joint Quantum Institute, University of Maryland, College Park, Maryland 20742, USA}
\author{Sankar Das Sarma}
\affiliation{Condensed Matter Theory Center, Department of Physics, University of Maryland, College Park}
\affiliation{Joint Quantum Institute, University of Maryland, College Park, Maryland 20742, USA}
\begin{abstract}
    The Gottesman-Knill theorem \cite{Gottesman98a} holds that operations from the Clifford group, when combined with preparation and detection of qubit states in the computational basis, are insufficient for universal quantum computation. Indeed, any measurement results in such a system could be reproduced within a local hidden variable theory, so that there is no need for a quantum mechanical explanation and therefore no possibility of quantum speedup \cite{Cuffaro13}. Unfortunately, Clifford operations are precisely the ones available through braiding and measurement in systems supporting non-Abelian Majorana zero modes, which are otherwise an excellent candidate for topologically protected quantum computation. In order to move beyond the classically simulable subspace, an additional phase gate is required. This phase gate allows the system to violate the Bell-like CHSH inequality that would constrain a local hidden variable theory. In this article, we both demonstrate the procedure for measuring Bell violations in Majorana systems and introduce a new type of phase gate for the already existing semiconductor-based Majorana wire systems. We conclude with an experimentally feasible schematic combining the two, which should potentially lead to the demonstration of Bell violation in a Majorana experiment in the near future. Our work also naturally leads to a well-defined platform for universal fault-tolerant quantum computation using Majorana zero modes, which we describe.
\end{abstract}
\maketitle
\section{Introduction}
    Universal quantum computation requires the operator to have the ability to produce any quantum state in the computational Hilbert space, including in particular those that violate the limits imposed on local hidden variable theories by the Bell inequality \cite{Bell64} or its variants such as the Clauser-Horne-Shimony-Holt (CHSH) inequality  \cite{Clauser69}. Even with the aid of measurement, however, a topological quantum computer based on the braiding of anyonic Majorana fermion zero modes (MZMs) cannot create such a state in a topologically protected manner \cite{Freedman03a, Nayak08}. This is intimately related to the fact that the braids and measurements of MZMs together form a representation of the Clifford group \cite{Nayak08,Alicea12a,Leijnse12,Beenakker2013a,Stanescu13,DasSarma15,Elliott15}, which is classically simulable \cite{Gottesman98a}. This well-known limitation of MZMs in carrying out universal quantum computation arises from the Ising (or $SU(2)_2$) nature of the corresponding topological quantum field theory (TQFT) which enables only $\pi/2$ rotations in the Hilbert space of the qubits through braiding. (The surface code implementation in superconducting
    qubits \cite{Fowler12}, which is turning out to be one of the most promising practical approaches to quantum computation at
    the present time \cite{Martinis14}, also suffers from the limitation of only supporting Clifford group operations in a
    a natural way.)

    While there are many theoretical proposals \cite{Bonderson10b,Mong14a,Stoudenmire15, Barkeshli13a, Barkeshli14} for going beyond MZMs (i.e. beyond the Ising anyon universality) which, in principle, could lead to universal topological quantum computation by utilizing richer  TQFT, e.g. $SU(2)_3$ or Fibonacci anyons, no such system has been experimentally demonstrated.  Furthermore, these richer systems enabling universal topological quantum computation often require extremely complicated braiding operations involving very high overhead in order to approximate Clifford group operations \cite{Kliuchnikov14}, which themselves are useful for quantum error correcting codes\cite{Terhal13}.  It is therefore of great importance to explore ideas which specifically utilize MZMs (with some additional operations) to carry out inherently quantum-mechanical tasks beyond the constraint of Gottesman-Knill theorem. In this paper, we propose what we believe to be the simplest nontrivial quantum demonstration feasible with MZMs, namely, a practical method for producing violations of the CHSH-Bell inequality, as well as a way of implementing the phase gate necessary to complement the existing Clifford operations and allow universal quantum computation with Majorana systems. The same ideas also lead directly to a practical platform that we introduce and discuss for performing universal quantum computation using Majorana wires.

    It is well known that the `protected' operations of braiding and measurement on MZMs (and also the surface code) become universal for quantum computation when supplemented with a single qubit phase gate of small enough angle \cite{Freedman06,Nayak08,Alicea12a,Leijnse12,Beenakker2013a,Stanescu13,DasSarma15,Elliott15}. In particular, the so-called $\pi/8$ phase gate (or $T$ gate) $e^{-i\pi/8}\ket{0}\bra{0}+e^{i\pi/8}\ket{1}\bra{1}$ is often named as part of a universal gate set. In part, this is because of the ``magic state distillation"\cite{Bravyi05} protocol that corrects errors in noisy $T$ gates through the use of Clifford gates and measurement. However, any phase gate with $\theta\neq n\pi/4$ with n integer is sufficient for universal quantum computation as long as it can be produced consistently. In fact, it would be useful for the reduction of overhead to be able to produce a phase gate of arbitrary angle, and there are now error correction algorithms designed with this in mind \cite{Duclos-Cianti14}.

    In this paper, we propose a measurement of violations of the CHSH-Bell inequalities as an important step in demonstrating not only the fundamental quantum physics and non-Abelian statistics of Majorana zero modes, but also the departure from the Clifford group that is necessary for universal quantum computation.  One great practical advantage of our proposal is that it builds on the existing proposals \cite{Sau10c,Sau11b,Alicea11,vanHeck12,Hyart13} for carrying out MZM braiding in semiconductor nanowire systems, which are currently being implemented in various laboratories on InSb and InAs nanowires \cite{Kouwenhoven15,Marcus15}. In contrast to the various exotic proposals for going beyond MZMs and $SU(2)_2$ TQFT \cite{Bonderson10b,Mong14a,Stoudenmire15, Barkeshli13a, Barkeshli14,Clarke13a,Lindner12,Cheng12}, our proposal does not necessitate building new experimental platforms, but uses the already existing and extensively experimentally studied semiconductor nanowire-based MZM platforms \cite{Alicea12a,Leijnse12,Beenakker2013a,Stanescu13,DasSarma15,Elliott15}.  In fact, our proposal may even be simpler than the existing proposals \cite{Sau10c,Sau11b,Alicea11,vanHeck12,Hyart13} for the nanowire MZM platform in the sense that no coupling needs to be fine-tuned between the various Majorana modes.

    Further, we introduce a new form of phase gate, improving on previous proposals for introducing an unprotected phase gate into a Majorana system in the sense that the gate we propose is adiabatic and therefore far less susceptible to timing errors than previous proposals \cite{Freedman06, Nayak08, Alicea12a,Beenakker2013a,DasSarma15, Bonderson10a, Clarke10, Hassler10, Hassler11}. This phase gate uses elements already present in the Hyart et al. proposal \cite{Hyart13} (semiconductor Majorana wires, superconductors, magnetic fields, and Josephson junctions), and so may be integrated into such a design without much additional overhead. Our new proposed phase gate should enable universal fault tolerant quantum computation using nanowire MZMs, and we introduce the precise platform for doing so in the laboratory.

    This paper is organized as follows: in Sec. \ref{Sec:CHSH} we discuss the CHSH type Bell inequalities in the context of MZM physics; in Sec. \ref{Sec:PhaseGate} we introduce the core idea of the new phase gate using a detailed analytical theory within a set of reasonable approximations; in Sec. \ref{Sec:Numerics} we provide numerical simulations to establish the validity of the theory; in Sec. \ref{Sec:Experiment} we provide the precise experimental protocols for realizing the phase gate and testing the Bell violation using MZMs; in Sec. \ref{Sec:Qubit} we introduce a new MZM nanowire-based physical platform for universal quantum computation; and we conclude in Sec. \ref{Sec:Outlook}.
\section{CHSH inequalities}\label{Sec:CHSH}

\begin{figure}[thb]
\includegraphics[trim=6cm 4.5cm 6cm 4cm, clip, width=\columnwidth]{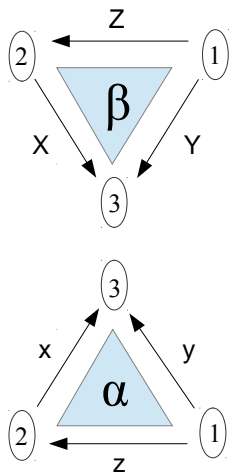}
\caption{This figure shows six Majorana zero modes arranged in two qubits consisting of three MZMs each, with pairs corresponding to the $x$, $y$ and $z$ axis of each qubit labeled. An arrow designates the definition of the corresponding Pauli operator in terms of the MZMs. For example, $\sigma_z=i\alpha_1\alpha_2$, $\sigma_y=i\alpha_1\alpha_3$.  }
\label{two spins}
\end{figure}

    To begin, we discuss the device independent aspects of our proposal, as the protocol we describe to test the CHSH inequalities is independent of the particular platform used to realize MZMs (and as such transcends the specific Majorana wire system of current experimental interest we focus on later in the paper). In order to test the inequality, we require a minimum of six Majorana zero modes, the ability to do a pairwise measurement of adjacent Majorana modes, and a phase gate implemented on (a particular) two of the Majorana modes. The procedure is as follows.

    First, we divide our six MZMs into two sets of three (See Fig.~\ref{two spins}). We label the Majorana fermion operators associated with the zero as modes $\alpha_i$ and $\beta_i$, where $i\in\{1,2,3\}$. The eigenvalues and commutation relations of the operators $i\alpha_i\alpha_j$ are such that we can make the identification
    \begin{equation}
      i\epsilon_{ijk}\alpha_i\alpha_j\equiv 2\sigma_k^\star
    \end{equation}
    where $\sigma_k$ is a Pauli matrix. In this way, we can identify the operators of three MZMs with those of a single spin-$\frac12$, such that the pairwise measurement of the state of two of the three MZMs corresponds to projective measurement along the $x$, $y$, or $z$ axis of the spin. We have labeled these psuedospin axes in Fig.~\ref{two spins}. We define Pauli matrices for the $\beta_i$ operators similarly and label the axes there with capital letters.

    The CHSH-Bell inequality \cite{Clauser69} now asserts that, in particular,
    \begin{equation}\label{bound-lhv}
      E(x,X)-E(x,Z)+E(z,X)+E(z,Z)\leq 2
    \end{equation}
    for a local hidden variable theory. Local in this case means local to the qubits, i.e. sets of three MZMs. The measurements we are making are necessarily \emph{non-local} in the individual MZMs themselves.

    That said, one may prepare a state that violates the CHSH inequality by first making initialization measurements that entangle the state of the two qubits. To this end, we begin by measuring the operators $i\alpha_1\beta_1$ and $i\alpha_2\beta_2$, projecting both into their $-1$ eigenstates.\footnote{If the `wrong' state is obtained, it may be corrected by measuring one MZM from the pair with one outside, then repeated until success \cite{Bonderson08b}.} Surprisingly, this alone is not enough to violate the inequality, but only to saturate it. In spin language, in the $z$-basis, we are in the state \mbox{$\frac{1}{\sqrt{2}}\ket{\uparrow\uparrow}+\ket{\downarrow\downarrow}$}. One may easily check that the expectation value of \mbox{$x\otimes X-x\otimes Z+z\otimes X+z\otimes Z$} for this state is indeed $2$. In fact, any set of measurements (or braiding) we do on the MZMs can only saturate the classical bound, never exceed it (thus not manifesting quantum entanglement properties, as is consistent with the Gottesmann-Knill theorem). To violate the CHSH inequality, we must add a phase gate (or equivalent) to the system. Applying this gate around the $y$-axis with an angle $\theta$, we find that
    \begin{equation}\label{Eq:CHSH-quant}
      E(x,X)-E(x,Z)+E(z,X)+E(z,Z)= 2\sqrt{2}\cos(2\theta-\pi/4)
    \end{equation}
    Note that the phase gate that is available from braiding alone has $\theta=\pi/4$, and therefore can only saturate the classical bound~(\ref{bound-lhv}). A more finely resolved phase gate than is available from braiding is necessary in order to violate the CHSH inequality. We discuss below how this can be done in a simple manner in order to directly observe quantum entanglement properties through the violation of CHSH-Bell inequality in an MZM-based platform.

\section{Phase Gate}\label{Sec:PhaseGate}
    There are now several proposals for introducing the necessary phase gate into a Majorana-based quantum computing scheme, including bringing the MZMs together in order to split the degeneracy for some time period\cite{Nayak08, Alicea12a,Beenakker2013a,DasSarma15}, using the topological properties of the system to create such a splitting non-locally\cite{Bonderson10a,Clarke10, Hassler10, Hassler11}, or transferring the quantum information to a different kind of qubit in order to perform the gate\cite{Bonderson11b}. Of these, the third requires a separate control scheme for the additional qubit, while the first two rely crucially on timing. In this Letter, we present a new type of phase gate whose elements are native to the Majorana wire platform, and which performs the phase rotation adiabatically, so that precise timing is not a concern.  The phase gate builds on previous proposals\cite{ Bonderson10a,Clarke10,Bonderson13b} that use the topological properties of the system in a non-topologically protected way in order to produce the gate \footnote{In many ways, what we describe is a practical implementation of the fine-tuned interferometry described in Ref.~(\onlinecite{Bonderson13b}).}. However, the fact that it is only a slight variation on already existing experimental setups \cite{Mourik12,Das12,Deng12,Rokhinson12,Finck12,Churchill13,Chang14} should presumably make our proposal easier to implement.

\begin{figure}
    \includegraphics[trim=4cm 6cm 10cm 6cm, clip, width=\columnwidth]{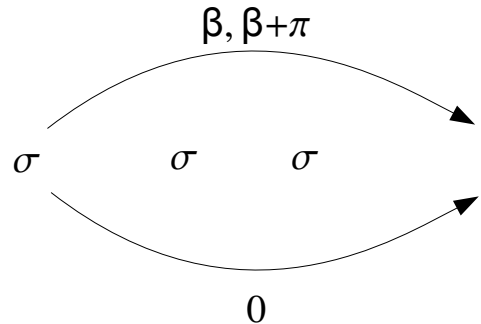}
    \caption{This figure shows the phase gained by an Ising anyon passing above or below a stationary Ising pair. The phase acquired is dependent on the combined state (qubit) stored by the two anyons. If the test particle passes entirely to one side of the pair, it acquires a topological phase of 0 or $\pi$ relative to passing on the other side in addition to the Abelian phase $\beta$. Quantum effects, whereby the test particle has an amplitude to pass on either side and these paths interfere, can lead to more general (though unprotected) phases. The role of the test particle in our proposal is played by a Josephson vortex, while that of the stationary pair is held by a Majorana wire placed on a superconducting island. }
    \label{fig-interferometer}
    \end{figure}

    We begin by considering the following thought experiment: Two Majorana zero modes (or Ising anyons) together form a two level system, which we may think of as the $\sigma_z$ component of a qubit. A third Ising anyon will pick up a topological component of phase $\pi(1-\sigma_z)/2$ upon going past this pair around the top, relative to the phase it picks up going around the bottom, in addition to any Abelian phase (See Fig.~(\ref{fig-interferometer})). If instead of giving this particle a classical trajectory, we allow it to behave quantum mechanically, then it now has some complex amplitude $A$ (or $B$) for going above (or below) the qubit pair as it moves from left to right. The total left to right amplitude is $A\sigma_z+B$. In the special case that the (Abelian) phases of $A$ and $B$ differ by $\beta=90^\circ$, the transmission probability is independent of the qubit state, and a phase gate
    \begin{equation}
    U(A,B)=\frac{1}{\sqrt{\abs{A}^2+\abs{B}^2}}(\abs{B}+i\abs{A}\sigma_z)=e^{i\arctan{\frac{\abs{A}}{\abs{B}}}\sigma_z}
    \end{equation}
    is applied to the qubit by the passage of the anyon.
    \begin{figure}
    \includegraphics[trim=5cm 5cm 7.4cm 5cm, clip, width=.9\columnwidth]{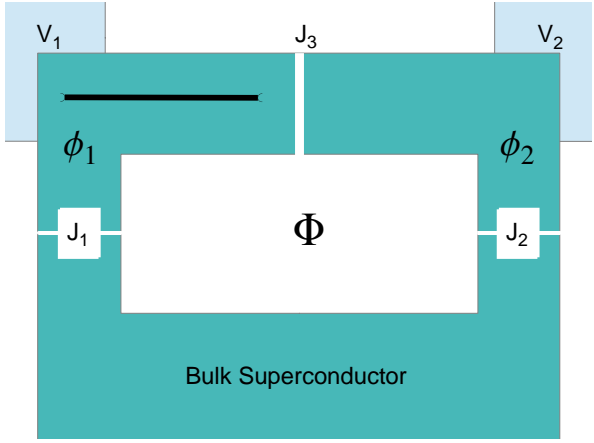}
    \caption{Proposed device for implementing a phase gate. A Majorana nanowire sits on the upper left of two superconducting islands connected to a bulk superconductor. Josephson Junctions $J_1$ and $J_2$ are adjustable, while a Josephson junction $J_3$ is strong and fixed. Gates of potential $V_1$ and $V_2$ are capacitively coupled to the superconducting islands. Operation of the phase gate is performed by ramping up the flux in the superconducting loop from $\Phi=0$ to $\Phi=2\pi$ while the strengths of couplings $J_1$ and $J_2$ are comparable to draw a Josephson vortex into the loop through two interfering paths, then ramping the flux back down to zero with $J_2\ll J_1$ to release the vortex deterministically through the right junction. }
    \label{Fig:PhaseGate}
    \end{figure}
    In order to realize this concept in a more physical (i.e. experimental) setting, we consider a ring of superconducting islands connected by three Josephson junctions (Fig.~\ref{Fig:PhaseGate}). Two of the Josephson junctions will be adjustable, while the third is assumed to be a much stronger link than the other two. One of these islands will hold a Majorana wire of the type described by Refs.~(\onlinecite{Lutchyn10,Oreg10, Sau10b}), which have already been the subject of experimental studies\cite{Mourik12,Das12,Deng12,Rokhinson12,Finck12,Churchill13,Chang14}. The endpoints of this wire act as Majorana zero modes, and allow that island to contain either an even or odd number of electrons with no energy penalty. We shall use the fermion parity ($q=\{0,1\}$) of this island as the axis of the qubit around which we perform our rotation. The role of the mobile Ising anyon in the above proposal may then be played by a magnetic flux vortex traveling through the Josephson junctions to enter the ring. The topological component of the phase picked up when a flux encircles the Majorana wire is again $\pi(1-\sigma_z)/2$, now arising from the Aharanov-Casher effect\cite{Aharanov84,Elion93,Konig06,Clarke10,Hassler10,Grosfeld11}. Finally, our setup includes a capacitive coupling to an adjustable gate voltage to one or both of the superconducting islands. This is represented in our model by the `gate charge' vector \mbox{$\vec{Q}=\left(\begin{array}{cc} C_{g1} V_{g1} & C_{2} V_{g2} \end{array}\right)$}
    where $C_{gi}$ and $V_{gi}$ are, respectively, the capacitance and voltage of the gates on each island. Changing
    $\vec{Q}$ allows us to adjust the relative (Abelian) phase $\beta$ acquired by the flux as it moves through one or the other of the weak Josephson links.
    \begin{figure}
    \includegraphics[width=.5\columnwidth]{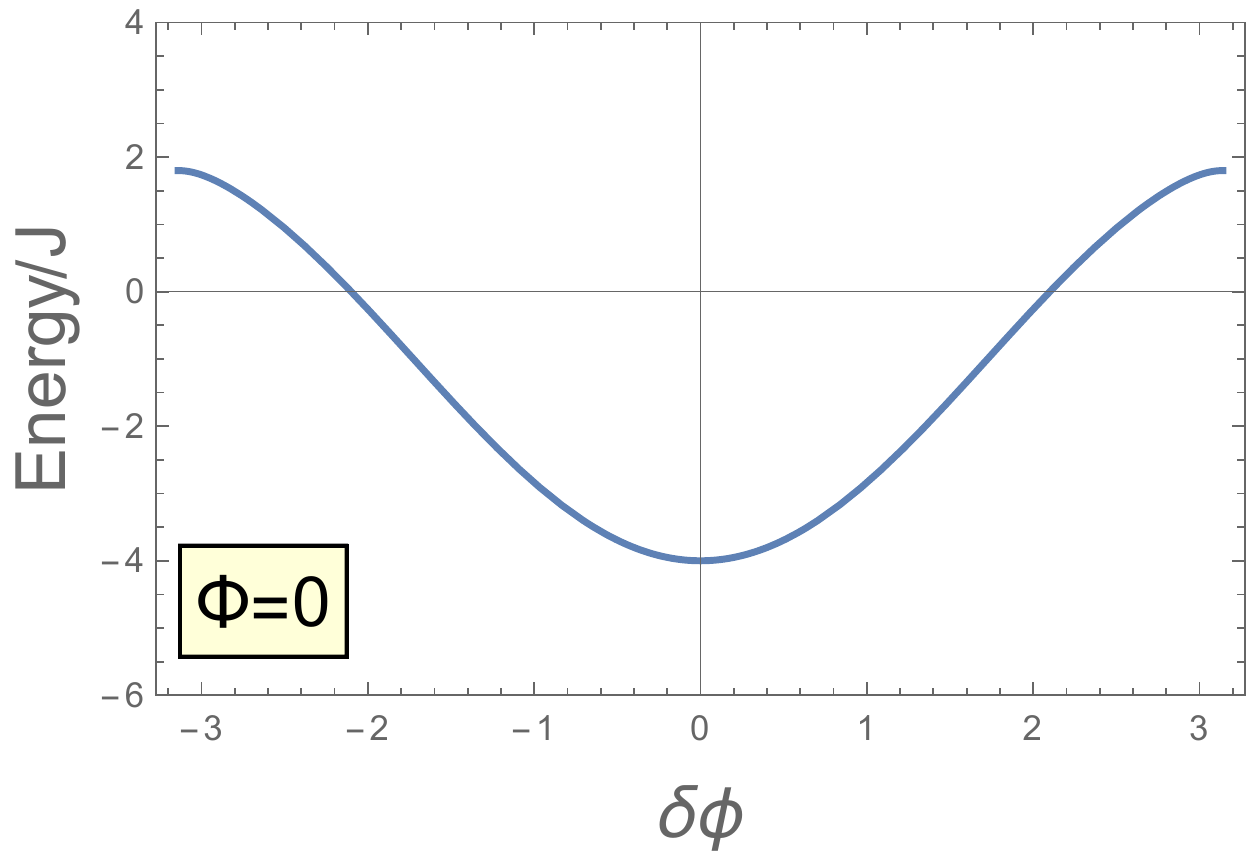}\includegraphics[width=.5\columnwidth]{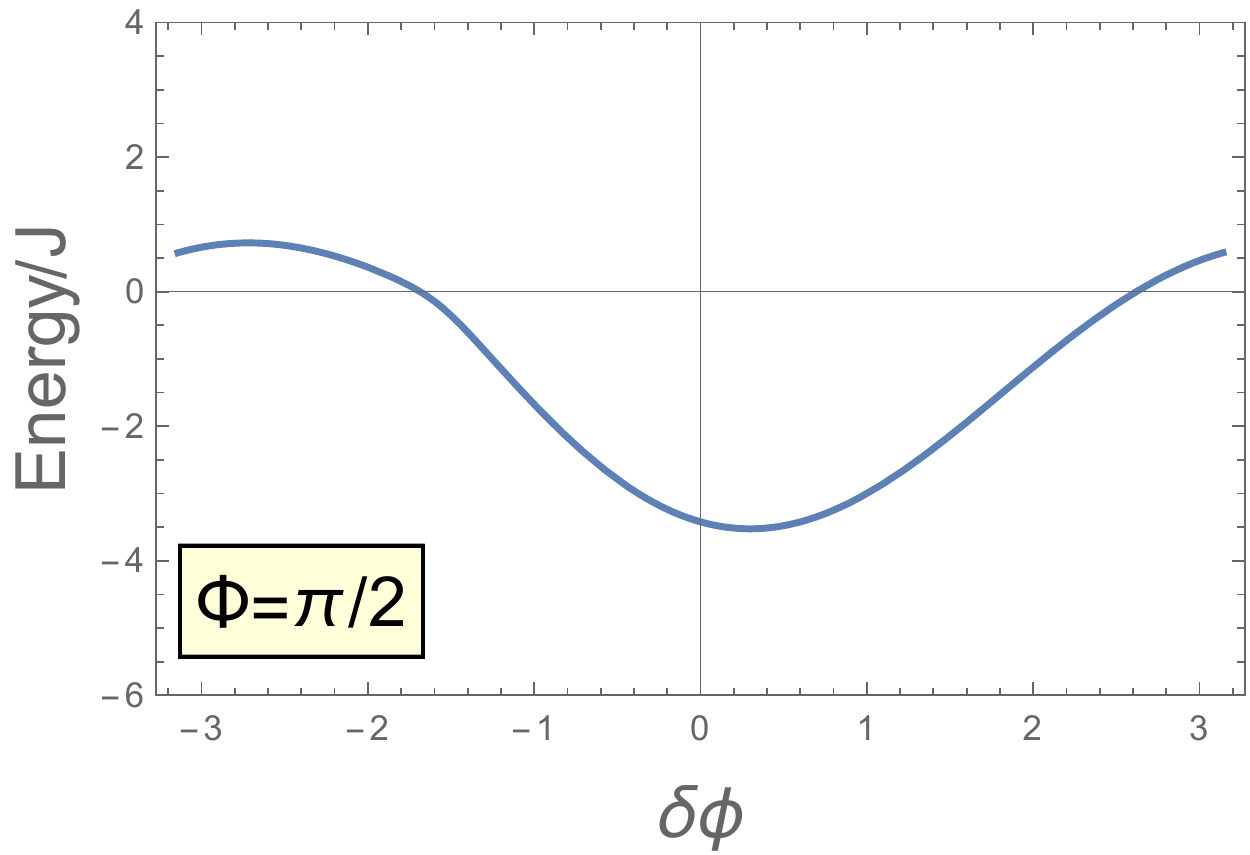}
    \includegraphics[width=.5\columnwidth]{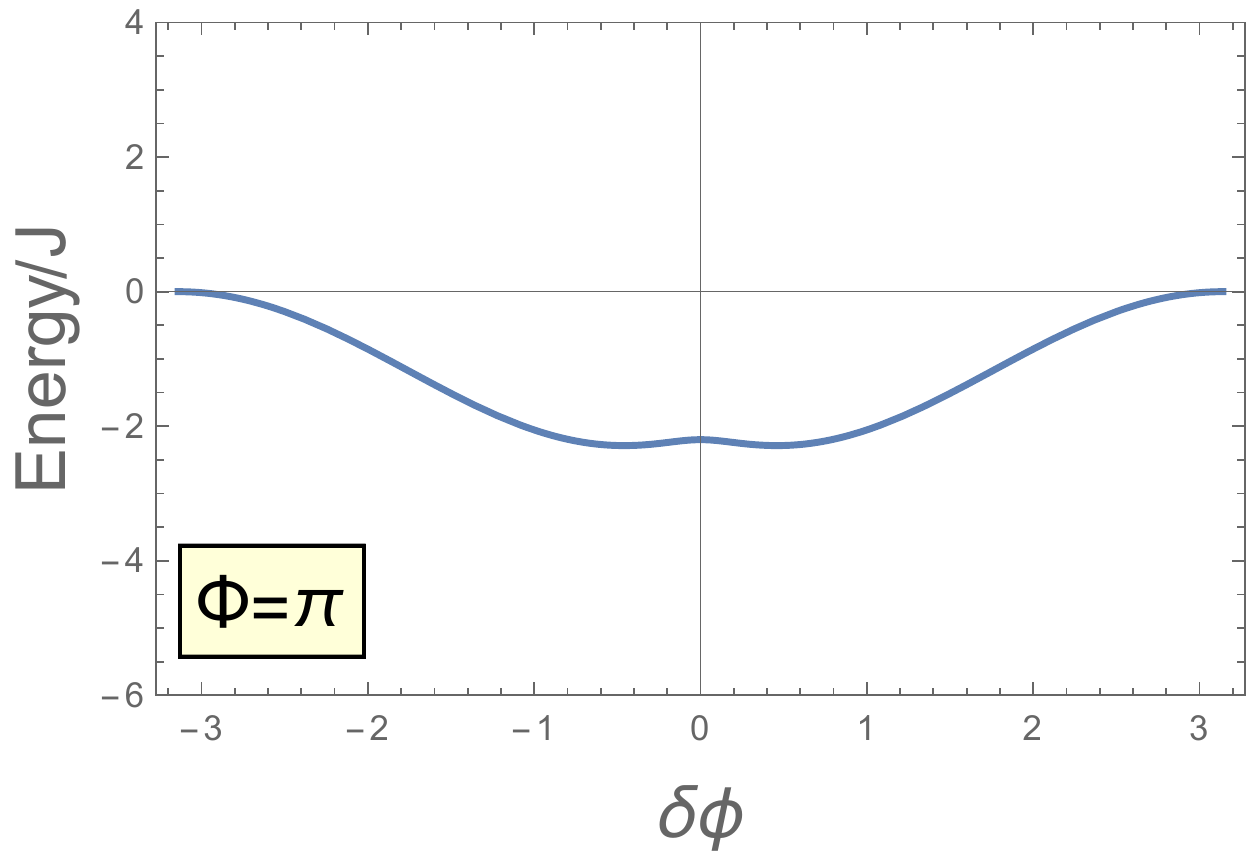}\includegraphics[width=.5\columnwidth]{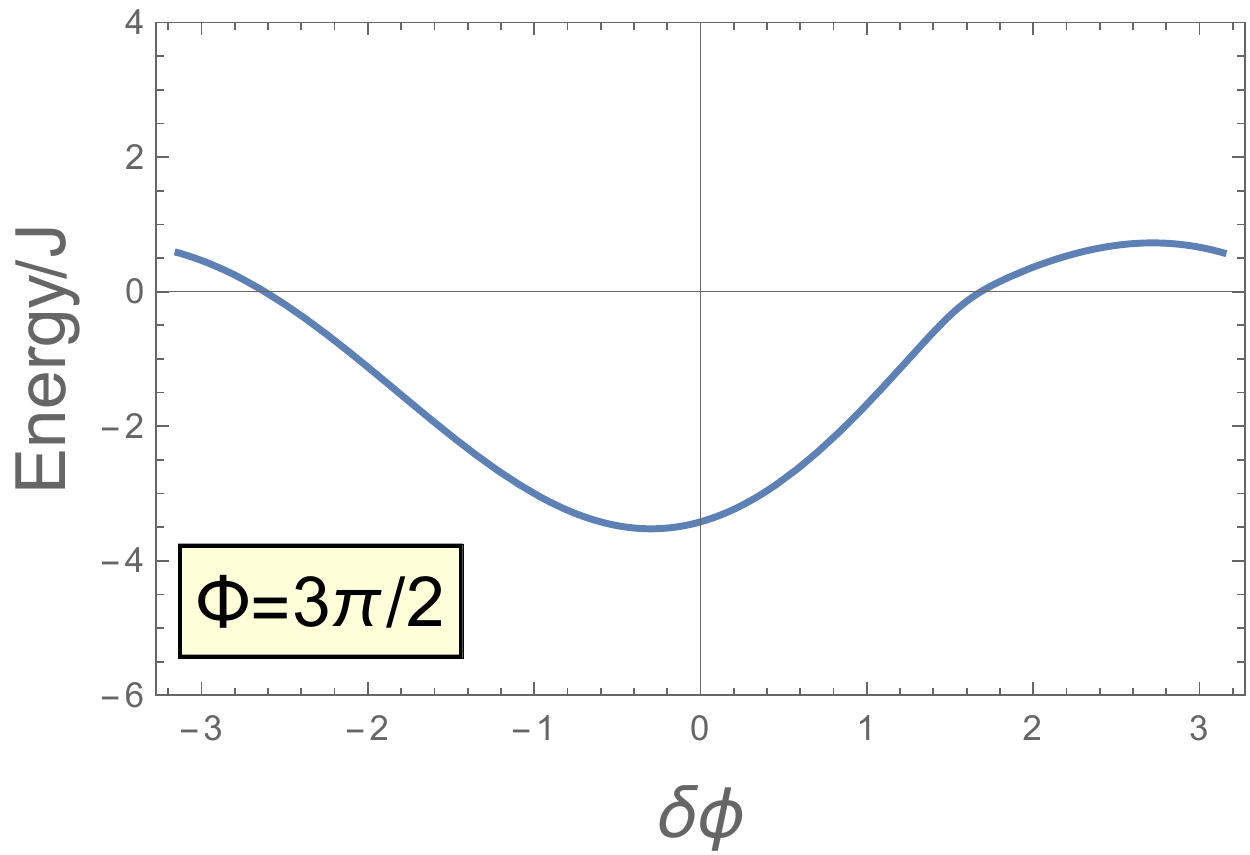}
    \caption{Potential energy vs $\delta\phi=\phi_1-\phi_2$ (minimized over $\bar{\phi}=(\phi_1+\phi_2)/2$) , plotted for $\alpha=2$, $\epsilon=.1$, and various values of $\Phi$. As $\Phi$ is tuned from $0$ to $2\pi$, a magnetic flux is drawn into the superconducting loop. Note the degeneracy at $\Phi=\pi$.}
    \label{fig:potentialplot}
    \end{figure}

    In order to implement our phase gate, we adjust the external magnetic field in order to slowly (adiabatically) increase the amount of magnetic flux running through the superconducting loop from $0$ to $2\pi$. (Note that there is no precise constraint on the exact timing of the flux threading process as long as it is adiabatic.) This will deterministically draw a Josephson vortex into the loop through one of the weak links, but crucially, does not measure which path that vortex takes. This is exactly the anyon interferometer we need to produce the phase gate\cite{Bonderson13b}.

    It remains to determine the phase that is produced based on the physical parameters of the system.  To do so, we begin with the Lagrangian
    \begin{equation}\label{Lagrangian}
        L=\frac{1}{2} \left(\frac{\Phi_0}{2\pi}\right)^2\dot{\vec{\phi}}C\dot{\vec{\phi}}^T +\frac{\Phi_0}{2\pi}\dot{\vec{\phi}}\cdot(\vec{Q}^T+e \vec{q}^T)
         -V(\phi_1,\phi_2),
    \end{equation}
    where \mbox{$\vec{\phi}=\left(\begin{array}{cc} \phi_1 & \phi_2\end{array}\right)$}, \mbox{$C=\left(\begin{array}{cc}
        C_1 & -C_3\\
        -C_3 & C_2
        \end{array}\right)$}, and
    \begin{equation}\label{eq:V}
      V(\phi_1,\phi_2)=-J_1\cos(\phi_1-\Phi/2)-J_2\cos(\phi_2+\Phi/2)-J_3\cos(\phi_1-\phi_2).
    \end{equation}
    Here \mbox{$\vec{q}=\left(\begin{array}{cc} q & 0\end{array}\right)$} is the fermion parity on the Majorana wire,
    \begin{figure}
    \includegraphics[width=.5\columnwidth]{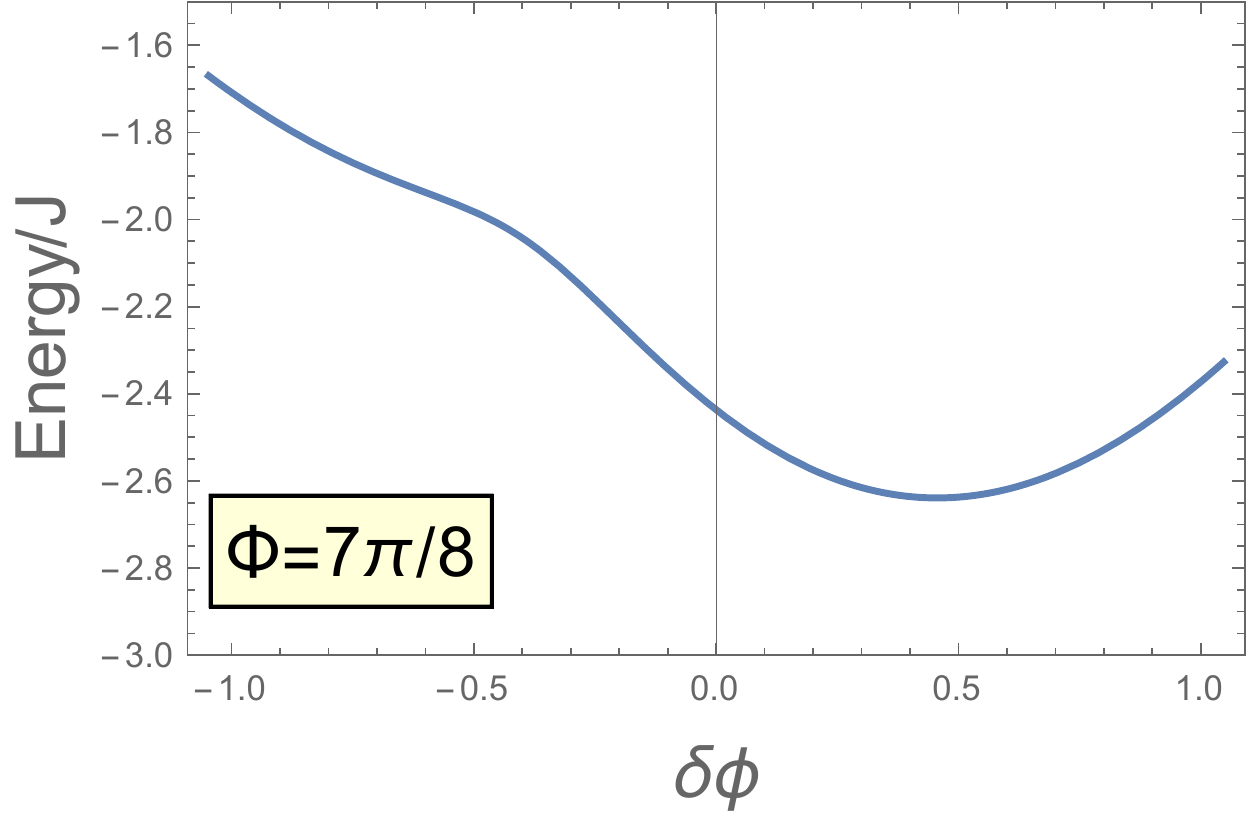}\includegraphics[width=.5\columnwidth]{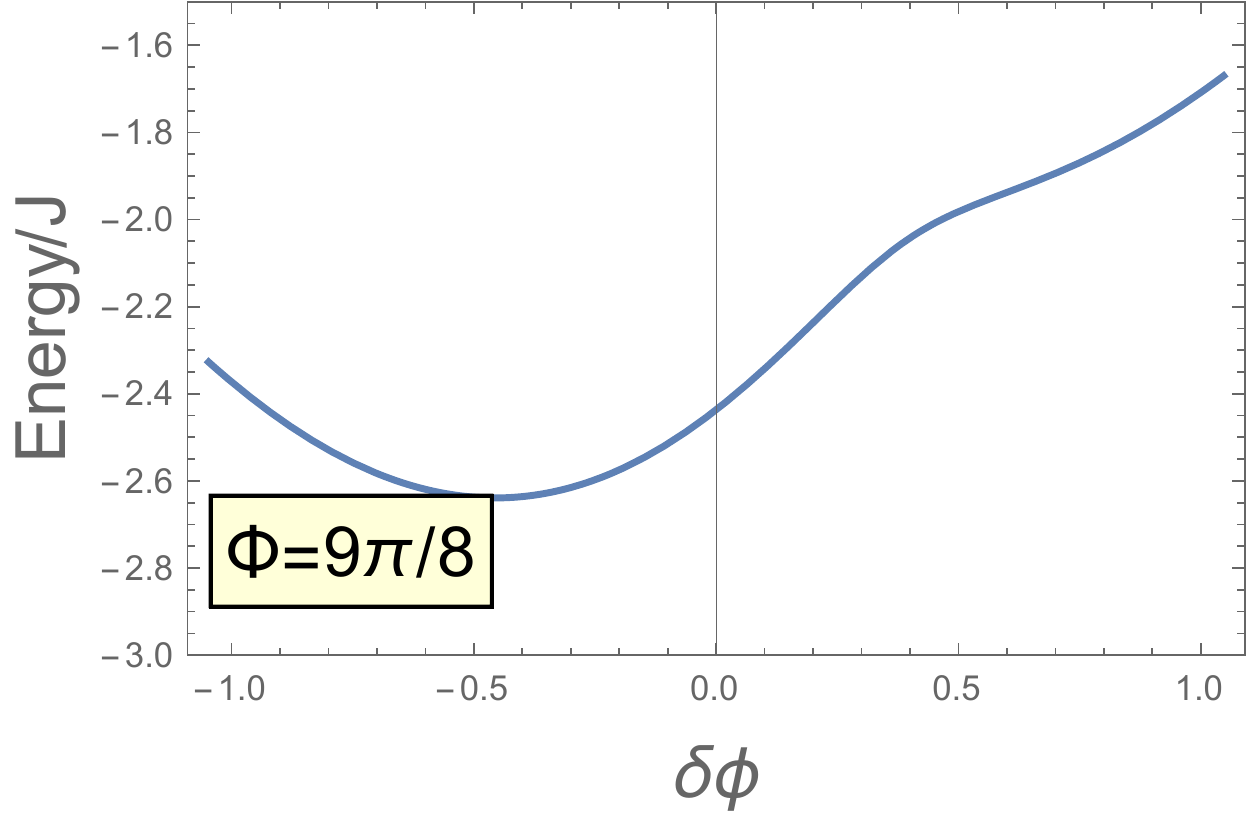}
    \includegraphics[width=\columnwidth]{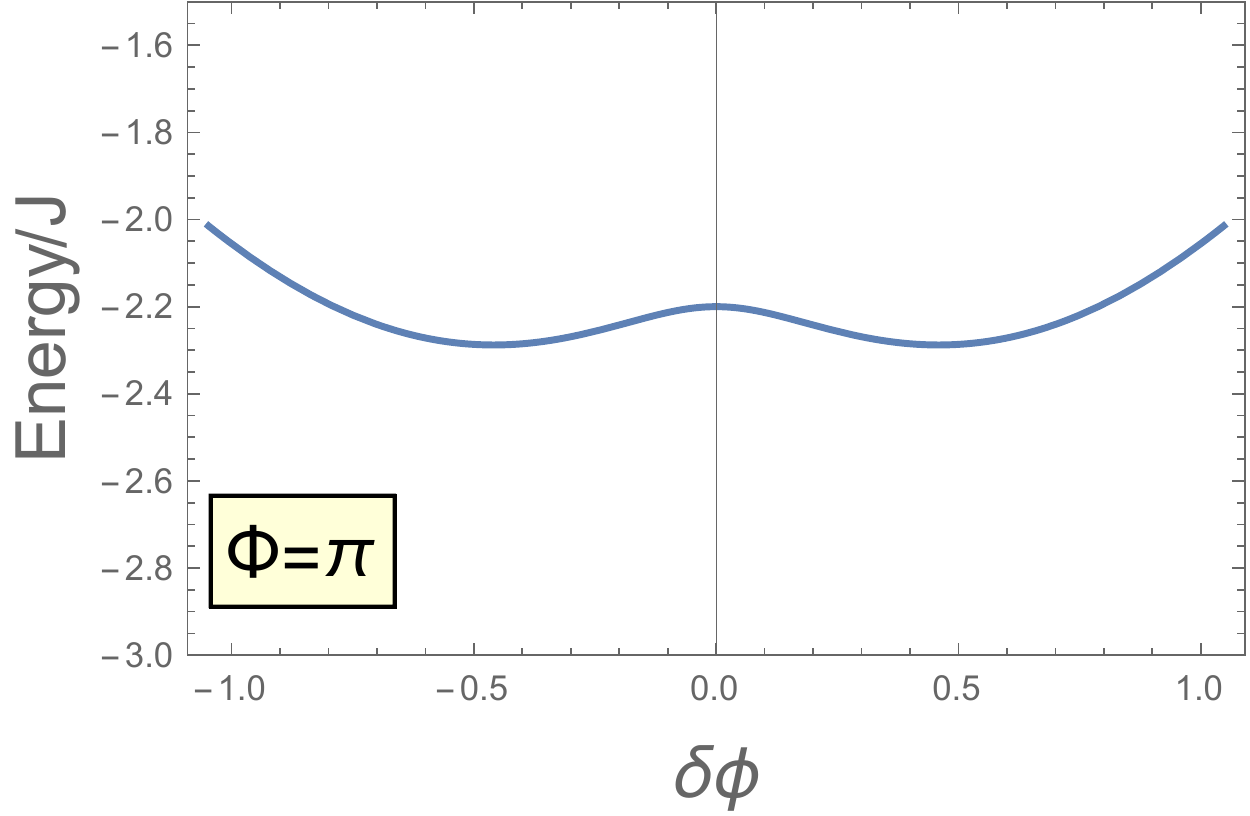}
    \caption{Potential energy vs $\delta\phi$ (minimized over $\bar{\phi}$) near the degeneracy point (plotted for $\alpha=2$, $\epsilon=.1$). Note that two inequivalent minima develop at $\Phi=\pi$, and an instanton event is required for the system to remain in the absolute minimum of energy as $\Phi$ is tuned past this point. This instanton may occur either with a forward or a backward jump in $\bar{\phi}$, and there is interference between the two paths.}
    \label{fig:potentialzoomed}
    \end{figure}
    the variable $\Phi$ is the flux through the superconducting ring, and $\dot{\phi}_i$ is the time derivative of the superconducting phase on island $i$.

    To run our phase gate, we will adiabatically increase the value of $\Phi$ from $0$ to $2\pi$ by applying an external magnetic field (Fig.~(\ref{fig:potentialplot})). We are most interested in the point $\Phi=\pi$, at which the system will need to cross a tunnel barrier to move from one degenerate minimum (the true minimum for $\Phi<\pi$) to the other (the true minimum for $\Phi>\pi$) (See Fig.~(\ref{fig:potentialzoomed})). We consider a system for which
    \begin{equation}\label{eq-2path}
        J_3(J_1+J_2)\geq J_1J_2\geq J_3 (J_1-J_2)\geq 0.
    \end{equation}
    In such a system the tunneling is well described by two interfering paths. Both paths will alter the phase difference $\delta\phi=\phi_1-\phi_2$ by the same amount. The paths differ by a full $2\pi$ winding of the average phase $\bar{\phi}=(\phi_1+\phi_2)/2$ of the superconducting islands.

    When $\Phi=\pi$, the degenerate minima of the potential $V$ occur at
    \begin{eqnarray}
      \cos(\delta\phi)&=&\frac{J_1^2+J_2^2}{2J_1J_2}-\frac{J_1J_2}{2J_3^2}\nonumber\\
      \tan(\bar{\phi})&=&\frac{J_1-J_2}{J_1+J_2}\cot(\delta\phi/2).
    \end{eqnarray}
    The value of the potential at these minima is
    \begin{equation}
      V_{\mathrm{min}}=J_3\frac{J_1^2+J_2^2}{2J_1J_2}-\frac{J_1J_2}{2J_3}.
    \end{equation}

    The classical equations of motion for the above Lagrangian (with $\Phi=\pi$) may be easily derived
    and rewritten as
     \begin{equation}
       \frac{\Phi_0^2}{8\pi^2J} \widehat{C}
        \left(\begin{array}{c}
        \ddot{\bar{\phi}} \\
        \delta\ddot{\phi}\end{array}\right)
        =\left(
       \begin{array}{cc}
         1 & \epsilon\\
        \epsilon & 1
        \end{array}\right)\left(\begin{array}{c} -\sin\bar{\phi}\sin\frac{\delta\phi}{2}\\ \cos\bar{\phi}\cos\frac{\delta\phi}{2}\end{array}\right)-\left(\begin{array}{c} 0\\ \alpha\sin(\delta\phi)\end{array}\right),
     \end{equation}
     where we have defined $J_1=(1+\epsilon)J$, $J_2=(1-\epsilon)J$,
     \begin{equation}
       \widehat{C}=\left(
       \begin{array}{cc}
         \bar{C} & \widetilde{C}\\
        \widetilde{C} & C_\delta
        \end{array}\right),
     \end{equation}
     and \begin{eqnarray}
       \bar{C}=C_1+C_2-2 C_3\nonumber\\
       \widetilde{C}=C_1-C_2\nonumber\\
       C_\delta=C_1+C_2+2C_3.\nonumber
     \end{eqnarray}
     We analyze these equations using an instanton approximation in the limit $\eta=\frac{1-\epsilon^2}{2\alpha}\ll 1$. Note that the condition~(\ref{eq-2path}) additionally requires that $\epsilon<\eta$.

     In this case, we can vastly simplify the equations of motion by expanding in orders of $\eta$:
     \begin{equation}\label{eq-expansion}
     \delta\phi=2\eta\cos\bar{\phi}-\frac{\Phi_0^2\eta\widetilde{C}}{4\pi^2 J}\ddot{\bar{\phi}}+\mathcal{O}(\eta^3).
     \end{equation}
     To bound the order of the corrections, we have used the fact that the first equation of motion implies that time derivatives scale as $\sqrt{\eta}$ because the $\bar{\phi}$ excursion for the instanton is not small.
     Next, we make use of energy conservation to gain the first integral of motion:
    \begin{equation}
      H=0=\frac{\Phi_0^2}{8\pi^2\bar{C}} (\bar{C}-\eta\widetilde{C}\sin\bar{\phi})^2\dot{\bar{\phi}}^2 +J\eta(\sin\bar{\phi}-\frac{\epsilon}{\eta})^2+\mathcal{O}(\eta^3).
    \end{equation}
    Note that for this equation to have a non-trivial solution for real $\phi_i$, we must propagate the system in imaginary time (hence the instanton solution).
    The total instanton action is therefore
    \begin{eqnarray}
      S&=&\int_{-i\infty}^{i\infty} \mathrm{d}t \sqrt{\frac{\Phi_0^2 J\eta}{2\pi^2 \bar{C}}} i\left(\bar{C}-\eta\widetilde{C}\sin\bar{\phi}\right)\left(\sin\bar{\phi}-\frac{\epsilon}{\eta}\right)\nonumber\\
      &+&\frac{\Phi_0}{2\pi}\left(Q_1+Q_2+e q\right)\dot{\bar{\phi}}+\frac{\Phi_0}{4\pi}\left(Q_1-Q_2+e q\right)\delta\dot{\phi}\nonumber\\
      &+&\mathcal{O}(\eta^3).
    \end{eqnarray}
    The last term provides a constant phase shift that is exactly canceled by the adiabatic phase coming from the change of the potential minimum for $\delta\phi$ as $\Phi$ is cycled from $0$ to $2\pi$ (unlike $\bar{\phi}$, $\delta\phi$ returns to zero after a full cycle). Likewise, we may ignore terms that are independent of the direction that $\bar{\phi}$ travels.
    In fact, we are only interested in the difference between the action of the paths with positive and negative $\dot{\bar{\phi}}$.
    The effective phase gate after adiabatic evolution of $\Phi$ is given by
    \begin{equation}\label{eq:pgate}
    U(q)=\exp\left(i\mathrm{Arg}(1+e^{i(S_+(q)-S_-(q))})\right),
    \end{equation}
    where $q$ is the qubit state and we can now calculate
    \begin{eqnarray}
      S_+-S_- &=&-i\Phi_0\sqrt{\frac{ J\eta}{2 \bar{C}}} \left(2\bar{C}\frac{\epsilon}{\eta}+\eta\widetilde{C}\right)+\mathcal{O}(\eta^{5/2})\nonumber\\
      &+&\Phi_0\left(Q_1+Q_2\right)+\pi q.
    \end{eqnarray}
    Note that the phase gate given by Eq.~(\ref{eq:pgate}) is gauge dependent. We have chosen the gauge in which tunneling a Josephson vortex through $J_1$ gives a $\pi$ phase difference between the two states of the qubit, while tunneling a vortex through $J_2$ does not measure the qubit charge. In order to get a gauge invariant quantity, we can reverse our procedure to release the vortex from the superconducting loop by ramping down the magnetic field, this time with $J_2$ tuned to $0$ so that the vortex has a guaranteed exit path. We shall not comment further on this second step of the procedure, and simply make the preceding (equivalent) gauge choice in what follows.

    The difference in instanton actions for the two entry paths takes the form $i(S_+-S_-)=i\pi q+i\beta-d$, where $\beta$ and $d$ are real numbers with $\beta=\Phi_0(Q_1+Q_2)$ and \mbox{$d\approx\Phi_0\sqrt{\frac{ J\eta}{2 \bar{C}}}  \left(2\bar{C}\frac{\epsilon}{\eta}+\eta\widetilde{C}\right)$}. In these terms, the phase accumulated between the two qubit states is given by
    \begin{equation}\label{Eq:pgate-phase}
    2\theta=\arg(\sinh(d)+i\sin(\beta))
    \end{equation}
    The $\pi/8$ phase gate appropriate to magic state distillation \cite{Bravyi05}, or for maximizing the violation of the CHSH inequality (see Eqs.~(\ref{bound-lhv}-\ref{Eq:CHSH-quant})), may be attained by choosing, \emph{e.g.}, $\beta=\pi/2$, $d=\asinh(1)$.

    One possible source of error is an induced splitting between the qubit states due to different rates of tunneling for the two qubit states near the instanton point $\Phi=\pi$, leading to a dynamic phase error in the qubit. Near the instanton point, the wave function is in a superposition between the left and right minima, and the energy of the lower state depends on the probability of the instanton event occurring. If this probability is different for different qubit states, the qubit will split.  The probability of the instanton event occurring for each state is proportional to
    \begin{equation}\label{eq:transition-prob}
        P(q)\propto \abs{1+e^{i(S_+(q)-S_-(q))}}=\abs{1+(-1)^q e^{-d+i\beta}}.
    \end{equation}
    This splitting puts a lower bound on how fast the phase gate should be performed, so as to minimize the accumulation of phase error. Note that if $\beta=\pi/2$, there is no splitting, as the probabilities are equal for the two qubit states. (This is also the condition that maximizes the controlled phase given by Eq.~(\ref{Eq:pgate-phase})). We expect the dynamic phase error to be minimized under this condition, an expectation which is (approximately) borne out by our numerical calculation.

    In the next section, we present the results of a numerical calculation that supports the analytical instanton analysis of this section.
\section{Numerical Simulation}\label{Sec:Numerics}

    In order to go beyond the instanton approximation detailed in the previous section, we performed numerical simulations of the Schrodinger equation associated with the Lagrangian~(\ref{Lagrangian}). The corresponding Hamiltonian for the system can be
    written as
    \begin{align}
    &H(\Phi)=E_C\sum_{j=1,2}(n_j-\frac{Q_j}{2e})^2-J_1\cos(\phi_1)\nonumber\\
    &-J_2\cos(\phi_2+\Phi)-J_3\cos(\phi_1-\phi_2),\label{eq:H}
    \end{align}
    where $E_C$ is the charging energy of each island (here we assume $C_1=C_2=2e^2/E_C$, $C_3=0$ for simplicity) and $n_j=-i\partial_{\phi_j}$ is the charge operator on each superconducting
    island. The Josephson energy part of the Hamiltonian, which is proportional to $J_{1,2,3}$, is identical to the potential used
    in the Lagrangian in Eq.~\ref{eq:V} up to a gauge transformation. For the numerical calculation, it is convenient to choose
    a gauge where the Hamiltonian is manifestly $2\pi$-periodic. Technically, there are three different such gauges where
    the flux enters across each of the junctions. As mentioned at the end of the last section, invariance with respect to the different
    gauge choices is guaranteed only when the Hamiltonian traces a closed loop where the flux $\Phi$ vanishes at the beginning
    and the end of the loop.

    In order to perform the simulation of the phase gate process, we divide the process of changing the flux $\Phi$ through the loop from 0 to $2\pi$ into a series of small time steps. At each step, we numerically solve the Schrodinger equation $H(\Phi)|\psi(\Phi)\rangle=E(\Phi)|\psi(\Phi)\rangle$ by expanding the wave-function in the eigenbasis $|n_1,n_2\rangle$ of the charge operators. The charging energy is diagonal in this basis and the Josephson energy terms are represented in terms of "hopping" terms such as $|n_1,n_2\rangle\langle n_1+1,n_2|$ etc. Choosing a large enough cutoff ($n_j\in [-15,15]$ turns out to be sufficient for our parameters), we can diagonalize the Hamiltonian matrix in the charge basis to obtain the ground state wave-function. Since the Hamiltonian is $2\pi$-periodic in the flux, the Berry phase can be computed from the expression
    \begin{equation}
    e^{i\theta_{\mathrm{Berry}}}\approx \prod_{n=1}^N\left\langle\psi\left(\frac{2\pi n}{N}\right)|\psi\left(\frac{2\pi(n+1)}{N}\right)\right\rangle,
    \end{equation}
    where  $N$ is the number of  steps into which we discretize the flux. Note that if we choose $N$ to be too small, the magnitude of the right hand side will be significantly less than unity, while as $N\rightarrow \infty$, the above approximation becomes exact \cite{Resta94}. The magnitude of the overlap at each step is thus an important diagnostic of the algorithm and should be near unity, serving as an important check on the accuracy of our simulation.

\begin{figure}
    \includegraphics[trim=0cm 6.5cm 0cm 7cm, clip, width=\columnwidth]{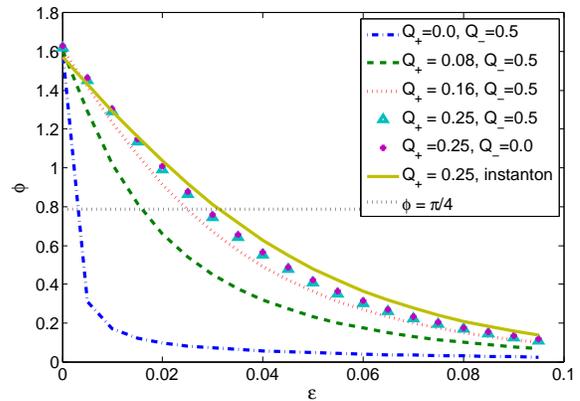}
    \caption{The relative phase acquired between states of the qubit after the phase gate is enacted, here plotted as a function of the Josephson junction asymmetry $\epsilon$ for a strong junction that is twice as strong as the weak junctions ($\alpha=2$). The charging energy $E_C$ has been chosen to be $0.4$ relative to the scale of the Josephson energy \mbox{$(J_1+ J_2)/2$}. $Q_+=(Q_1+Q_2)/$ and $Q_-=Q_1-Q_2$ are expressed in units of the Cooper pair charge $2e$. Note that a relatively small junction asymmetry of $\epsilon\leq .1$ can tune the gate through a large range of phases. }
    \label{fig:phase}
    \end{figure}

    Fig.~\ref{fig:phase} shows the relative phase between the two qubit states acquired through the adiabatic evolution. Note that the `magic' phase $\pi/4$ \cite{Bravyi05} can be attained either by adjusting the gate voltages to change $Q_i$, or by adjusting the imbalance $\epsilon$ in the Josephson couplings. We can compare these results to the prediction of the instanton approximation using our calculated $d\approx 4\pi\sqrt{\frac{ \alpha J}{E_C(1-\epsilon^2)}}\epsilon $.

    The qubit state $q$ is encoded in the Hamiltonian through  a shift of the gate charge $Q_1\rightarrow Q_1+q$.
    The qubit phase generated from the phase gate is calculated by calculating the difference of Berry phases $\theta_{Berry}(q=1)-\theta_{Berry}(q=0)$ acquired as the flux $\Phi$ is changed by $2\pi$. To maintain adiabaticity, the flux must be swept at
    a rate that is small compared to the first excitation gap $E_{gap}$ above the ground state $|\psi(\Phi)\rangle$ of the Hamiltonian in Eq.~\ref{eq:H}. Such a slow sweep rate leads to a dynamical contribution to the qubit phase that is given by
    \begin{align}
    &\theta_{dyn}=\int d\Phi \frac{E(\Phi,q=1)-E(\Phi,q=0)}{\dot{\Phi}}.
    \end{align}
    To keep this error small, the sweep rate $\dot{\Phi}$ must be kept larger than the
    energy difference, \emph{i.e.} \mbox{$|E(\Phi,q=1)-E(\Phi,q=0)|\ll\dot{\Phi}$}. At the same time,
    adiabaticity requires \mbox{$\dot{\Phi}\ll E_{gap}$}. Thus, the dynamical range (\emph{i.e.} the range of sweep rates)
    over which this gate can operate, is proportional to
    \begin{align}
    \zeta=\frac{E_{gap}}{|E(\Phi,q=1)-E(\Phi,q=0)|}.
    \end{align}
    The inverse of the dynamical range $\zeta^{-1}$ also quantifies the contribution of the dynamical phase to the
    systematic error in the gate.

    \begin{figure}
    \includegraphics[trim=0cm 6.5cm 0cm 7cm, clip, width=\columnwidth]{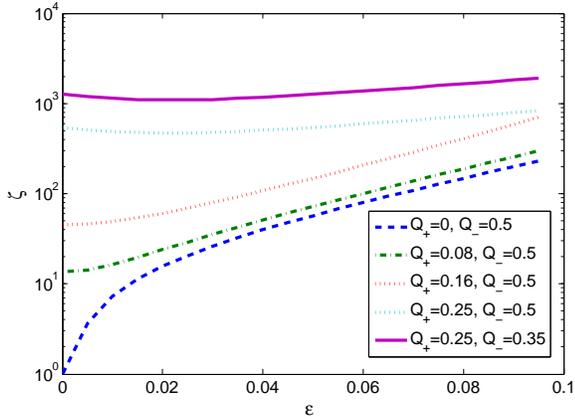}
    \caption{The dynamical range of the phase gate as a function of $\epsilon$ for a set of values of $Q_+=(Q_1+Q_2)/$ and $Q_-=Q_1-Q_2$ (expressed in units of $2e$). The best dynamic range is found for $Q_1+Q_2=1/4$, corresponding to a phase of $\beta=\pi/2$ in Eq.~(\ref{eq:transition-prob}). Here the dynamic phase error and the minimal gap differ by three orders of magnitude, allowing the gate to function for a significant range of ramping times.}
    \label{fig:dynamicrange}
    \end{figure}
    \begin{figure}
    \includegraphics[trim=0cm 6.5cm 0cm 7cm, clip, width=\columnwidth]{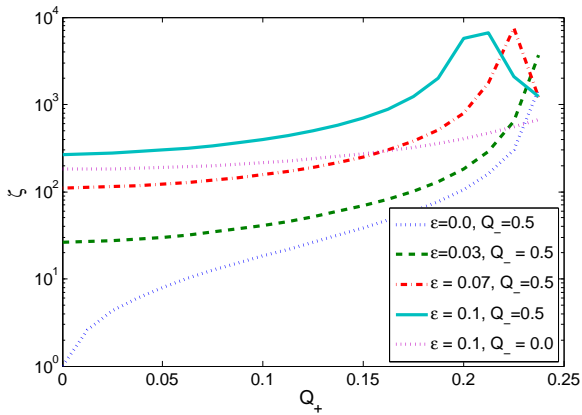}
    \caption{The dynamical range of the phase gate as a function of $Q_+=(Q_1+Q_2)/$ (units of $2e$) for a set of values of $\epsilon$ and $Q_-=Q_1-Q_2$ (expressed in units of $2e$). Note the resonances in dynamic range as the qubit states are tuned to degeneracy.}
    \label{fig:dynamicrange2}
    \end{figure}
     While at the lowest order instanton approximation, the energy $E(\Phi,q)$ is independent of q,  as seen from
    the numerical results in Fig.~\ref{fig:dynamicrange}, higher order instanton corrections
    lead to energy splittings that are a finite fraction $\zeta^{-1}>0$ of the gap. This is apparent from Fig.~\ref{fig:dynamicrange}, since
    in the ideal case $\zeta$ would be infinite. However, it is also clear that the leading order contribution to $\zeta$
    can be minimized by choosing the total gate charge near $Q_1+Q_2=0.25$.  This is expected if the major contribution to the qubit splitting comes from the instanton contribution described in Sec.~\ref{Sec:PhaseGate}. $Q_1+Q_2=0.25$ corresponds to $\beta=\pi/2$ in Eq.~\ref{eq:transition-prob}. Fig.~\ref{fig:dynamicrange2} shows that the resonance in the dynamic range (corresponding to the degeneracy point for the qubit states) does not always occur exactly at $Q_1+Q_2=0.25$. Higher order corrections to the instanton calculation will detect the asymmetry in the system, leading to dependence of the resonance on $Q_1-Q_2$ and $\epsilon$. Nevertheless, it is evident that a dynamic range of two to three orders of magnitude is achievable over a broad range of parameter space, enabling a rather unconstrained experimental implementation of the phase gate without undue fine tuning.  Based on these results, we suggest in the next section a precise experimental scheme to implement our proposed phase gate as well as to verify the CHSH-Bell inequality mentioned in the title of our paper.

    \begin{figure}
    \includegraphics[trim=6cm 5.5cm 6cm 5.5cm, clip, width=\columnwidth]{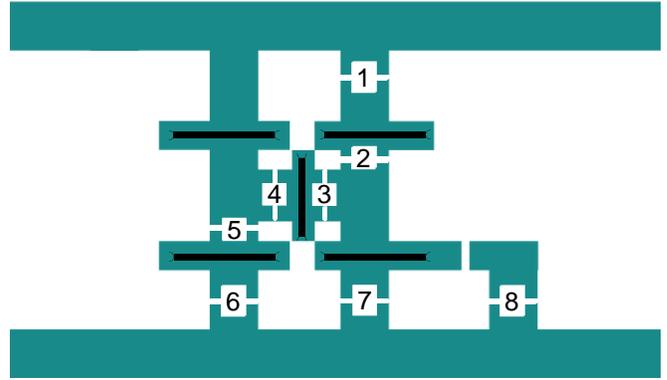}
    \caption{Experimental Design for CHSH Measurement. The system consists of several superconducting islands placed within a microwave resonator. Five Majorana nanowires are placed among the islands in such a way as to produce six Majorana zero modes, one at each endpoint along the outer edge, and one at each intersection where three wires meet \cite{Hyart13}. The top three Majoranas form one qubit, and the bottom three another. Eight adjustable Josephson junctions (and one strong fixed junction) couple the islands to each other and to the Bus (top) and Phase Ground (bottom) of the microwave resonator. Of these junctions, those labeled 1-6 need only have `on' (strongly coupled) and `off' (very weakly coupled) settings. Junctions 7 and 8 are used to implement the phase gate as described in the main text.}
    \label{Fig:Resonator}
    \end{figure}
   \begin{figure}
   \includegraphics[trim=0cm 3cm 0cm 3cm, clip, width=.9\columnwidth]{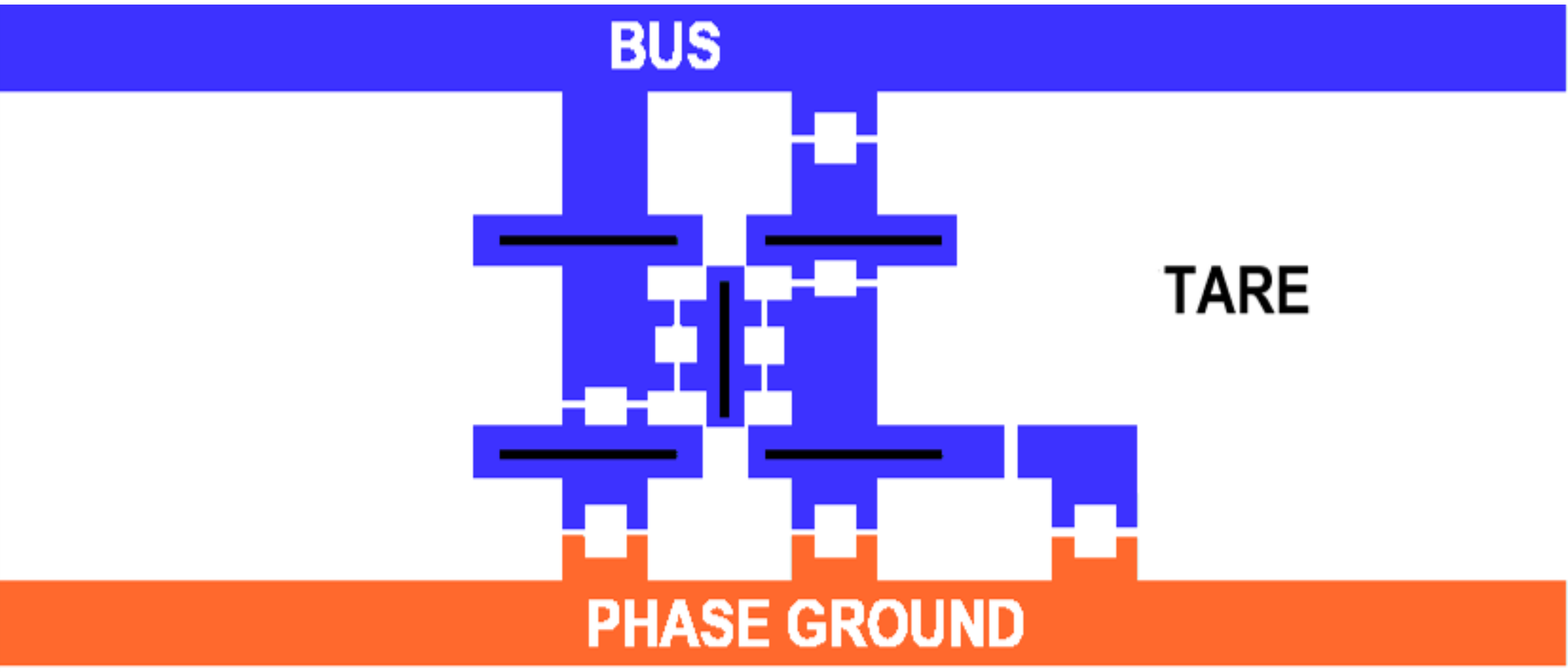}
   \includegraphics[trim=0cm 3cm 0cm 3cm, clip, width=.9\columnwidth]{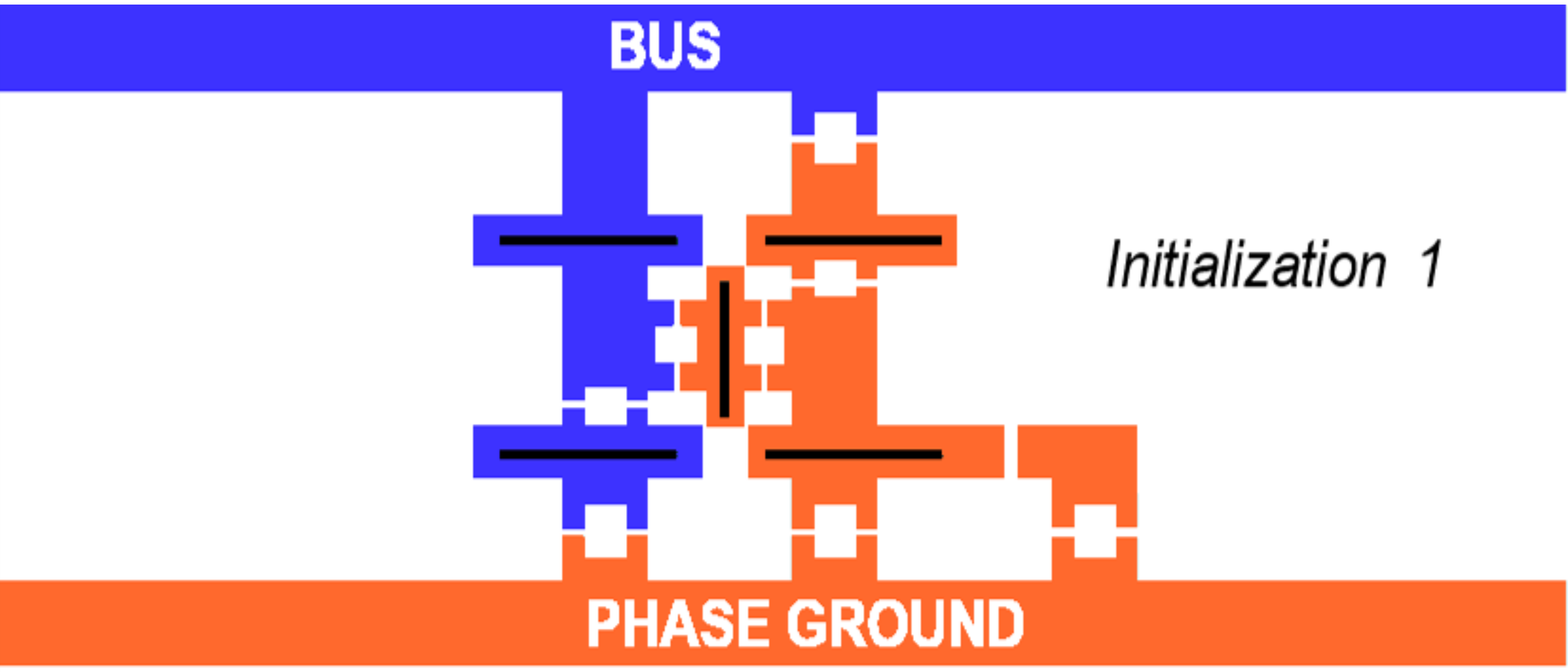}
   \includegraphics[trim=0cm 3cm 0cm 3cm, clip, width=.9\columnwidth]{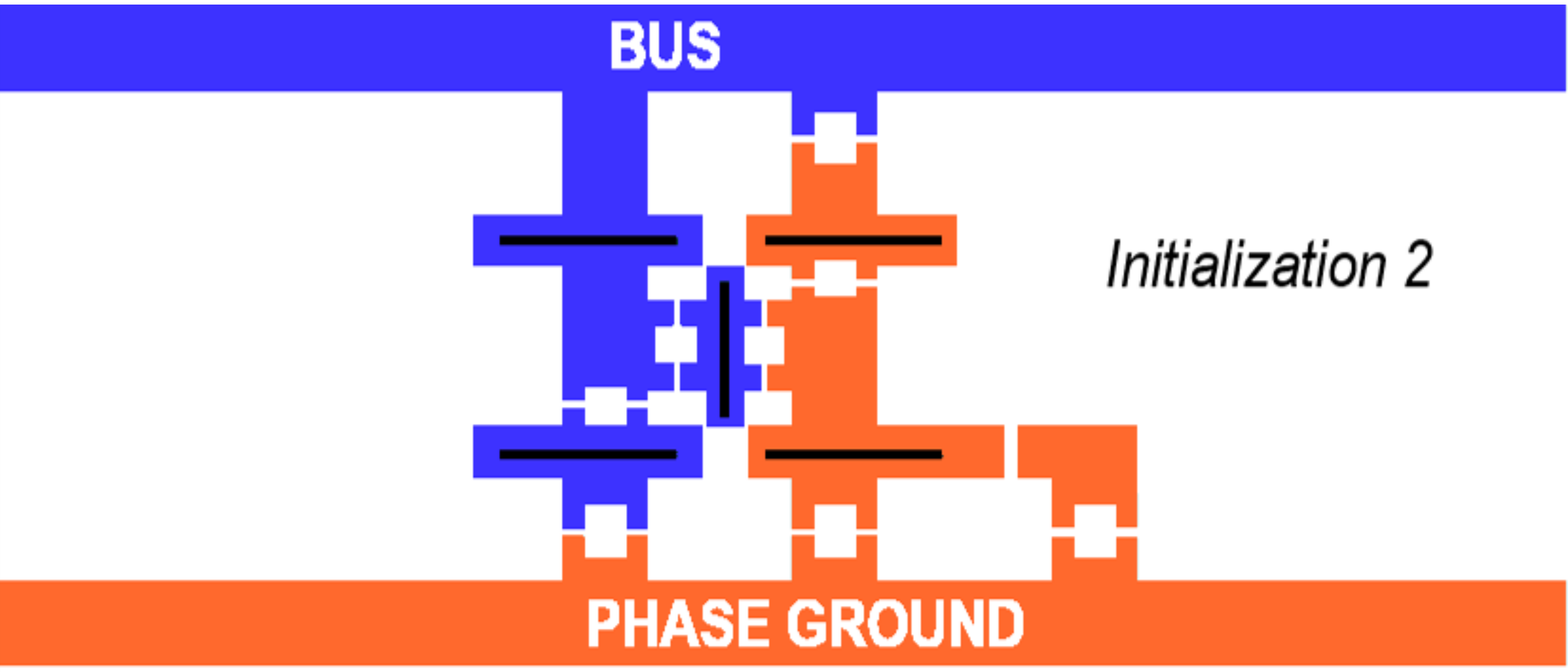}
    \caption{(Color Online) Initialization process. The Blue (Dark Gray) and Orange (Light Gray) superconducting islands are strongly Josephson coupled to the Bus and Phase Ground respectively. (Top) Coupling all islands to the Bus gives a measurement of the overall fermion parity of the system. This allows one to equivalently measure the parity of a qubit by coupling the constituent Majorana modes to the Bus or the Phase Ground. (Middle,Bottom) The two qubits in this device may be put into the superposition $\frac{1}{\sqrt{2}}(\ket{00}+\ket{11})$ by making two measurements; each measurement includes one MZM from each qubit. (Along with the tare measurement, this will also effectively measure the parity of the central island).   }
    \label{Fig:Initialization}
    \end{figure}
   \begin{figure}
   \includegraphics[trim=0cm 3cm 0cm 3cm, clip, width=.9\columnwidth]{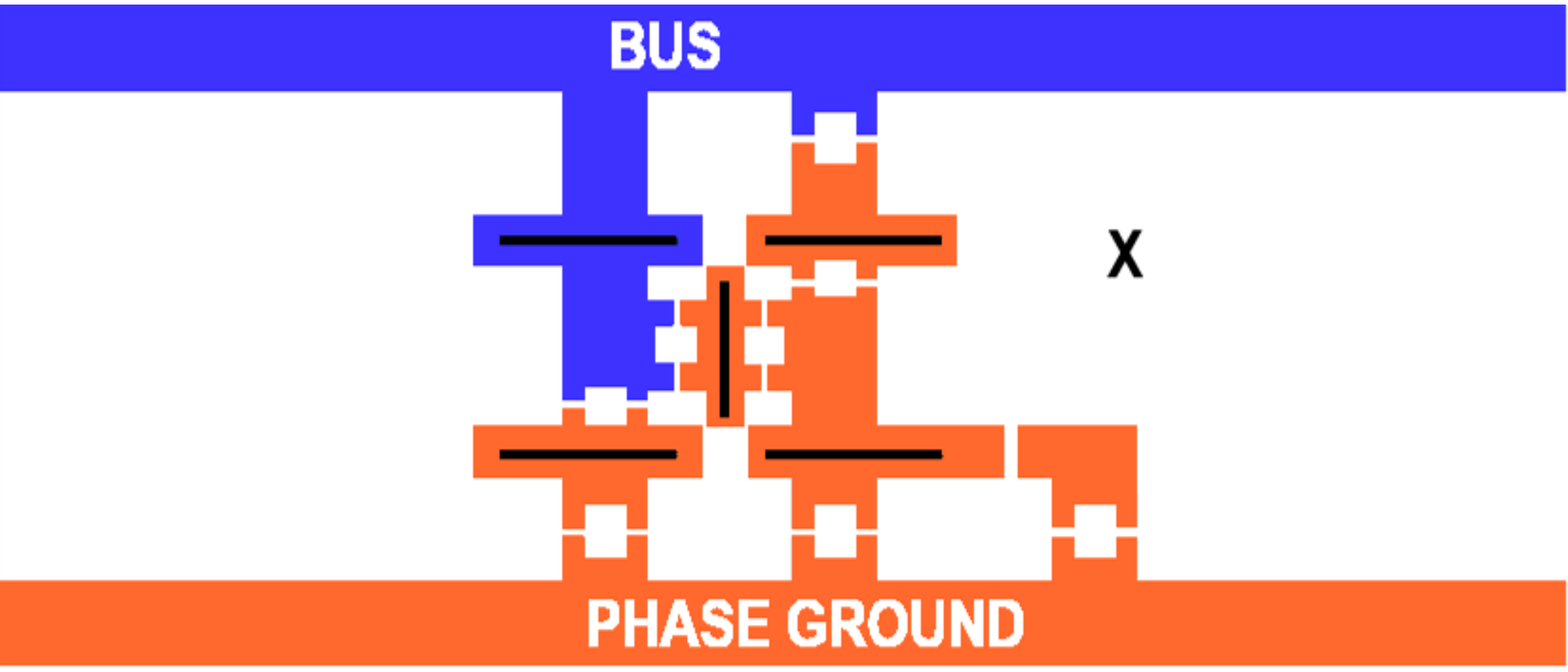}
   \includegraphics[trim=0cm 3cm 0cm 3cm, clip, width=.9\columnwidth]{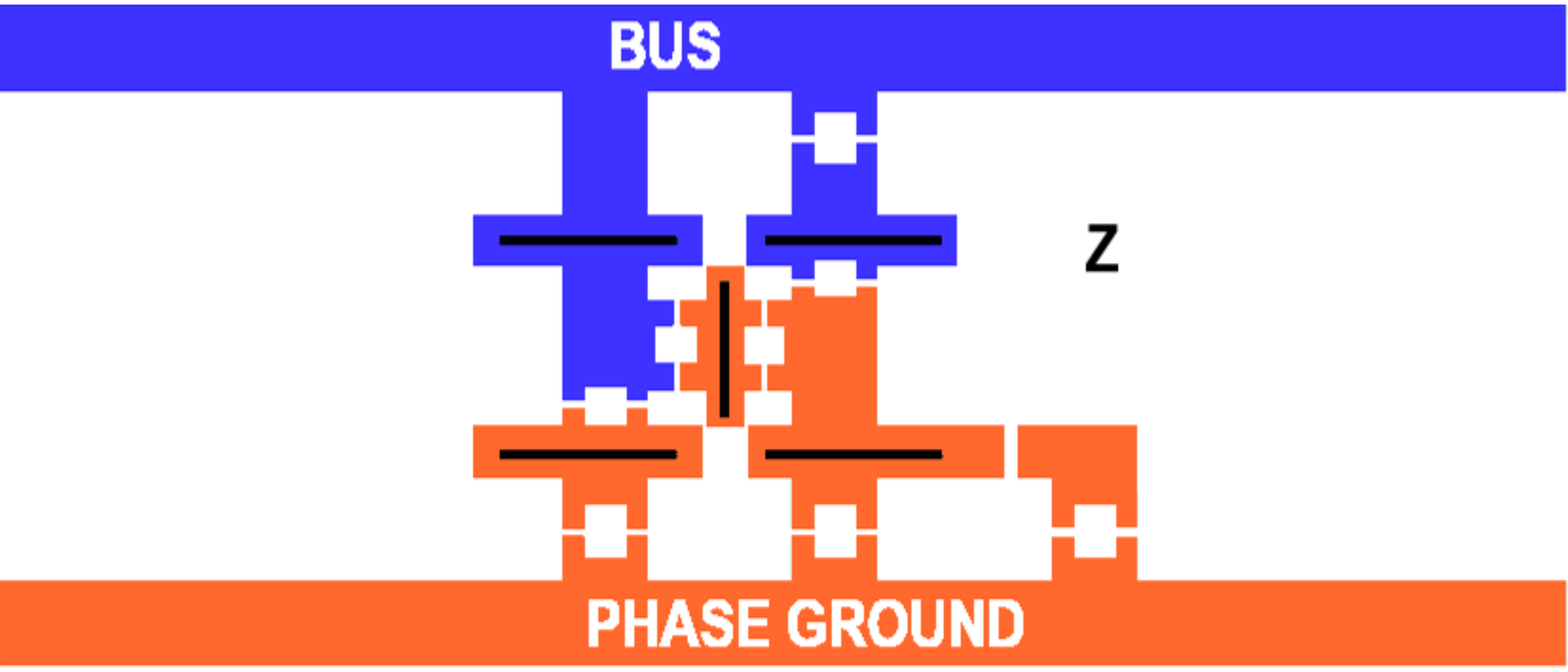}
    \caption{(Color Online) Measurement of the $X$ and $Z$ projections of the upper qubit. The parity of a set of MZMs is measured by Josephson coupling the corresponding superconducting islands strongly to the Bus, while coupling the remaining islands to the Phase Ground.   }
    \label{Fig:UpperQubit}
    \end{figure}
   \begin{figure}
   \includegraphics[trim=0cm 3cm 0cm 3cm, clip, width=.9\columnwidth]{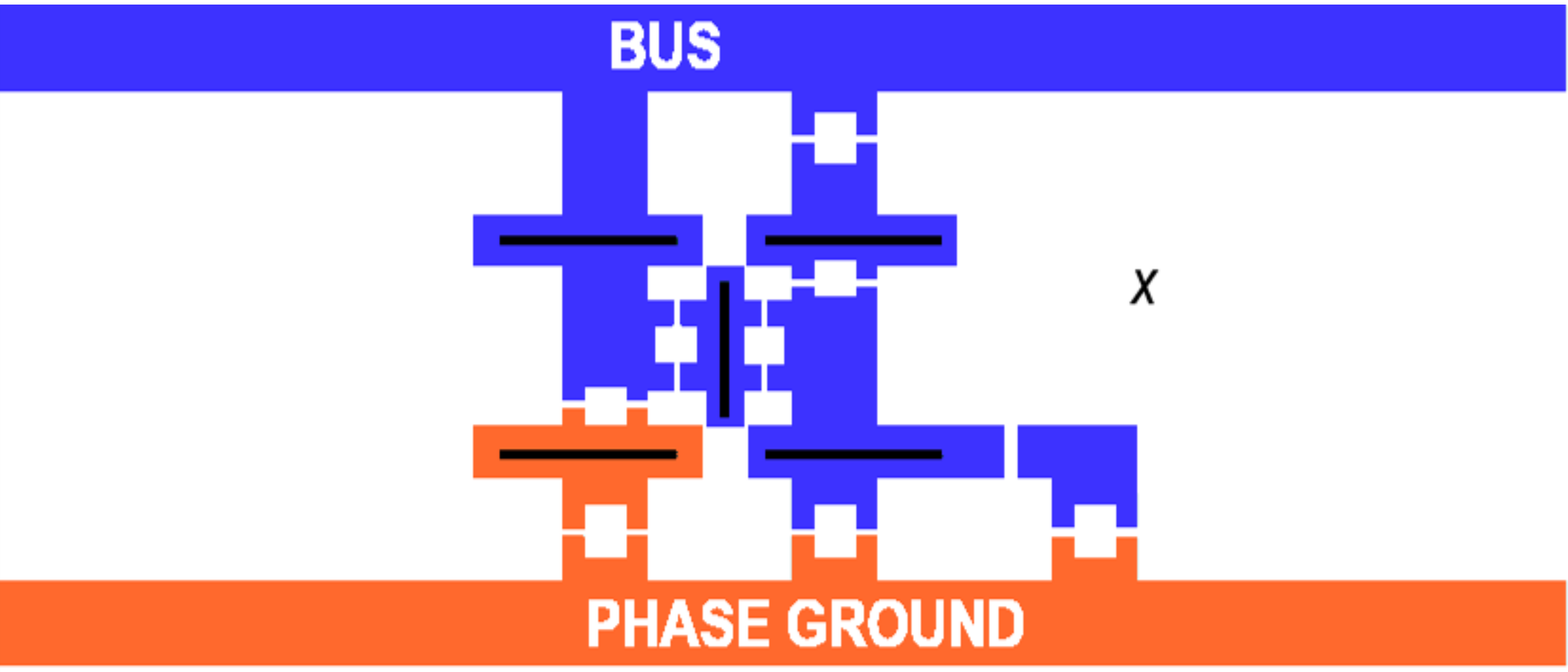}
   \includegraphics[trim=0cm 3cm 0cm 3cm, clip, width=.9\columnwidth]{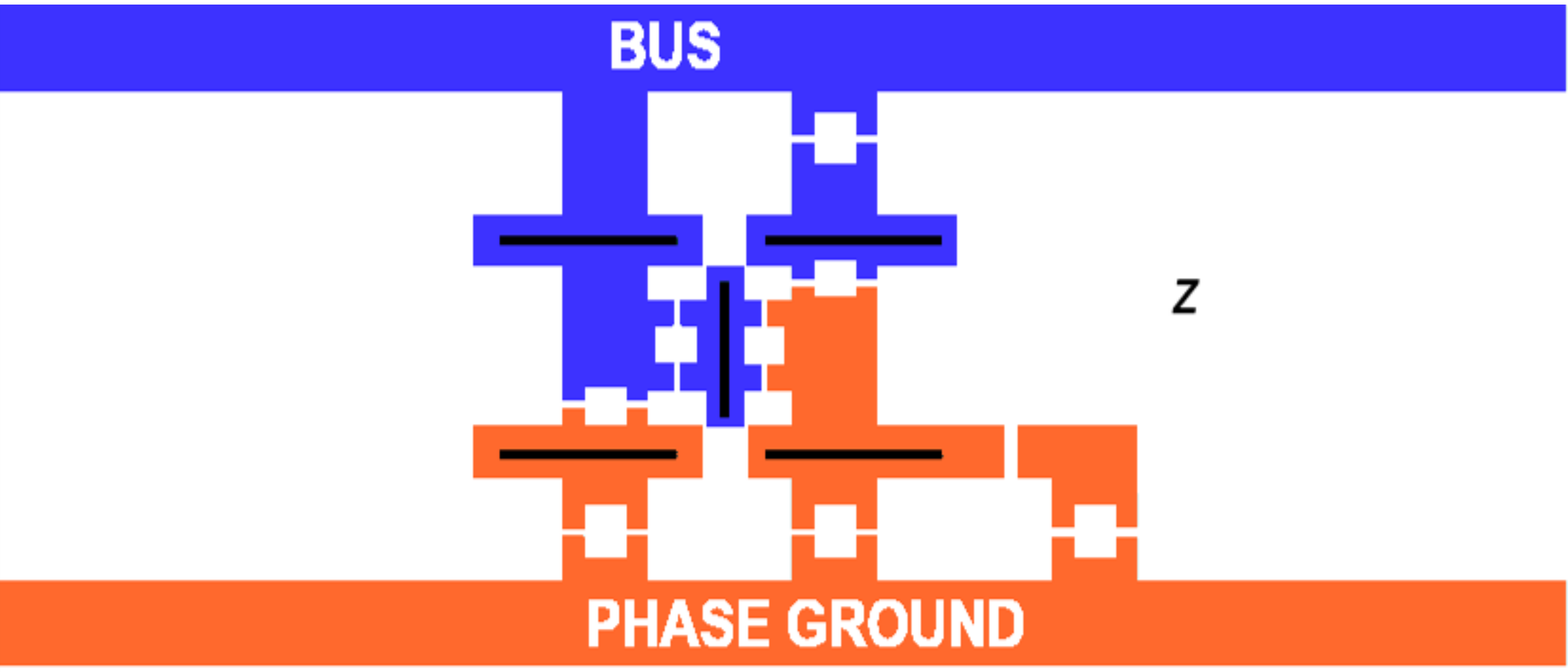}
    \caption{(Color Online) Measurement of the $x$ and $z$ projections of the lower qubit. The tare measurement (See Fig.~\ref{Fig:Initialization}) allows one to relate the parity of the section coupled to the Phase Ground to the parity of the section coupled to the Bus.   }
    \label{Fig:LowerQubit}
    \end{figure}

\section{Experimental Proposal}\label{Sec:Experiment}
    In order to implement and test this phase gate within a physical system, we turn to the measurement scheme proposed by Hassler et al. \cite{Hassler11} (and used extensively by Hyart et al. \cite{Hyart13}) in which the parity of a pair of Majorana fermions is read out through a superconducting charge qubit in a transmission line resonator (a `transmon' \cite{Koch07}). Using the arrangement shown in Fig.~\ref{Fig:Resonator}, we can implement all of the steps of the CHSH-Bell test without ever needing to physically braid any of the Majoranas.  This is in line with the so-called `Measurement Only' schemes for topological quantum computation \cite{Bonderson08b,Bonderson08c}. In our proposed experiment, each of the islands containing Majorana wires is attached through a series of adjustable Josephson junctions to the larger superconducting region on either the top (Bus) or bottom (Phase Ground). We assume that the coupling between the three Majoranas at each of the trijunctions is strong compared with the charging energy of the islands, and is comparable to the Josephson energy at junction $J_8$ during the performance of the phase gate. Due to this large coupling, the Majorana at the left end of the wire in Fig.~\ref{Fig:PhaseGate} is effectively replaced by the zero mode of the trijunction.

    Measurement of the resonance frequency of this system when placed within a microwave resonator can resolve whether the total parity of islands connected to the Bus is even or odd \cite{Koch07,Hassler10,Hyart13}. Again, the trijunction coupling is assumed to be large compared with the measurement scale so that the trijunctions are treated as effectively single MZMs. The $X$ and $Z$ components of the upper qubit may be measured by connecting the islands containing the corresponding MZMs strongly to the Bus (while connecting all other islands to the Phase Ground, see Fig.~\ref{Fig:UpperQubit}). Similarly, measurements of the lower qubit may be made by connecting the corresponding islands strongly to the Phase Ground (Fig.~\ref{Fig:LowerQubit}). Because the resonator measurement determines the total parity connected to the Bus, the parity of the islands connected to the Phase Ground may be inferred once the overall parity of the Majorana system is measured. We label this a `Tare' measurement (See Fig.~\ref{Fig:Initialization}).

    Together, Figs.~\ref{Fig:Initialization}-\ref{Fig:LowerQubit} show the sets of islands coupled to Bus and Phase Ground corresponding to each of the measurements necessary for a test of the CHSH inequality. In most cases, we need only two settings for our Josephson junctions, strong connection $E_J\gg E_C$ (On), and very weak connection $E_J\ll E_C$ (Off). This switching can be controlled, e.g., by a threading half a flux between two strong Josephson junctions\cite{Hyart13}. The two Josephson junctions in the lower right of Fig.~\ref{Fig:Resonator}, labeled $7$ and $8$, adjust the phase gate in the manner described above, acting as control parameters for $J_1$ and $J_2$. A strong Josephson coupling $J_3$ is assumed between the two islands in the lower right. While the phase gate is being implemented, Junctions $2$ and $3$ should be off, allowing flux to pass freely between the island containing the lower right Majorana wire (`$y$') and the remaining wires. The loop in the lower right now acts as the superconducting loop of Fig.~\ref{Fig:PhaseGate} for implementing the phase gate around the $y$-axis. The remaining Josephson junctions should be on, so that no measurement path is open in the resonator system and all other islands have the superconducting phase inherited from the Phase Ground. In this case, the Josephson energy of the lower trijunction acts to renormalize the coupling $J_1$ in the phase gate design of Fig.~\ref{Fig:PhaseGate}.

    Figs. \ref{Fig:Resonator}-\ref{Fig:LowerQubit} provide the schematic and the protocol for the experimental platform as well as the necessary measurements for the phase gate and CHSH violation being introduced in this work. They also provide all the tools necessary to conduct an independent test of the fidelity of the phase gate without any alteration of the device, e.g. by measuring $x$ before and after a phase rotation around $y$.
\section{Qubit design for universal quantum computation}\label{Sec:Qubit}
\begin{figure}
    \includegraphics[trim=0cm 3cm 0cm 3cm, clip, width=.45\columnwidth]{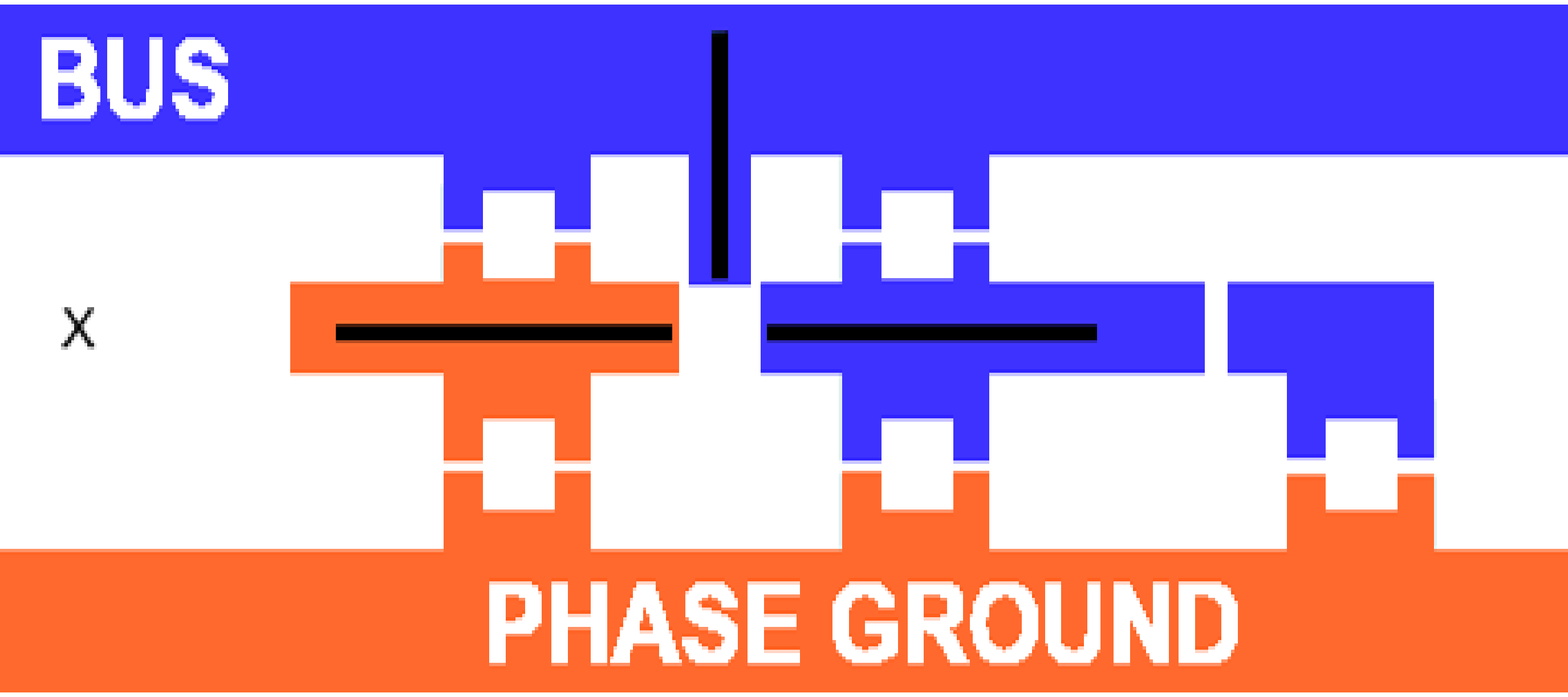} \quad \quad \includegraphics[trim=0cm 3cm 0cm 3cm, clip, width=.45\columnwidth]{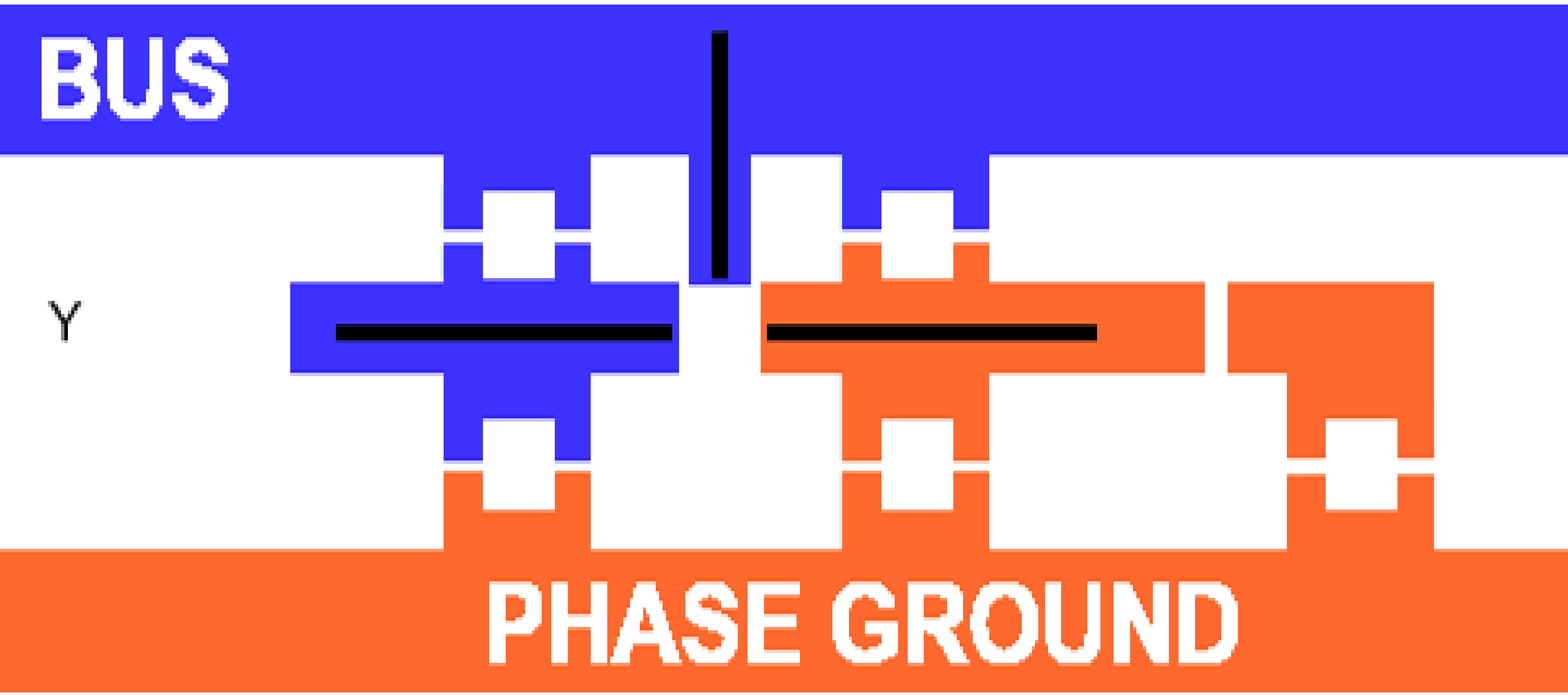}
    \includegraphics[trim=0cm 3cm 0cm 3cm, clip, width=.45\columnwidth]{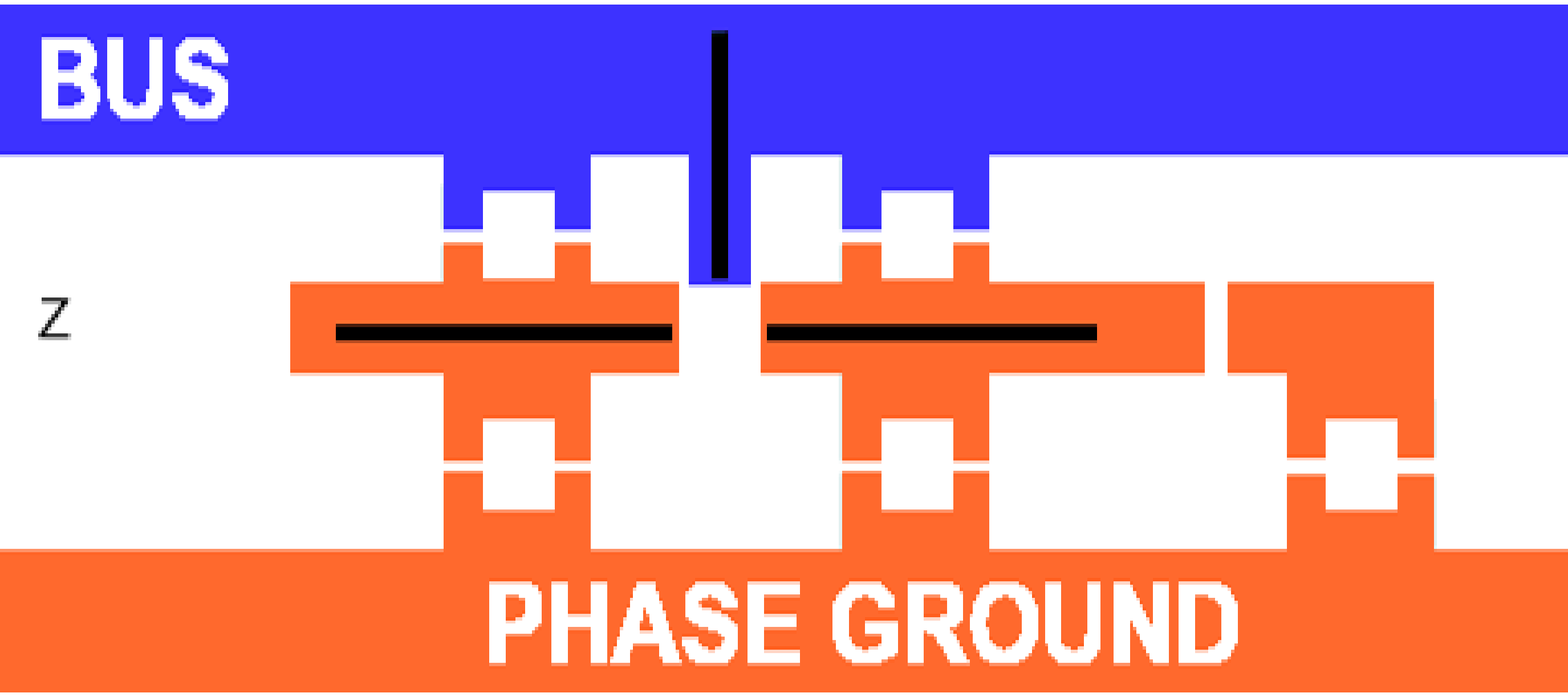} \quad \quad \includegraphics[trim=0cm 3cm 0cm 3cm, clip, width=.45\columnwidth]{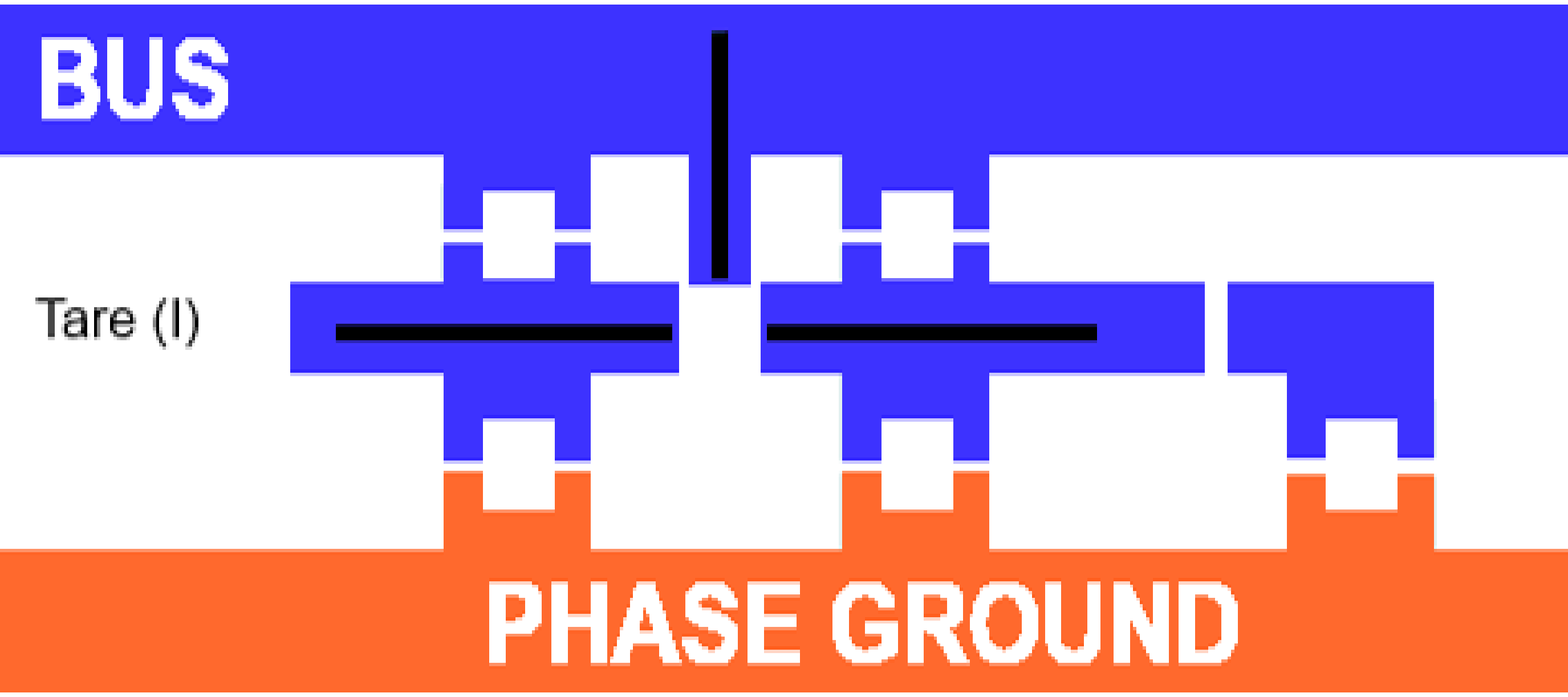}
    \includegraphics[trim=0cm 8cm 0cm 8cm, clip, width=.9\columnwidth]{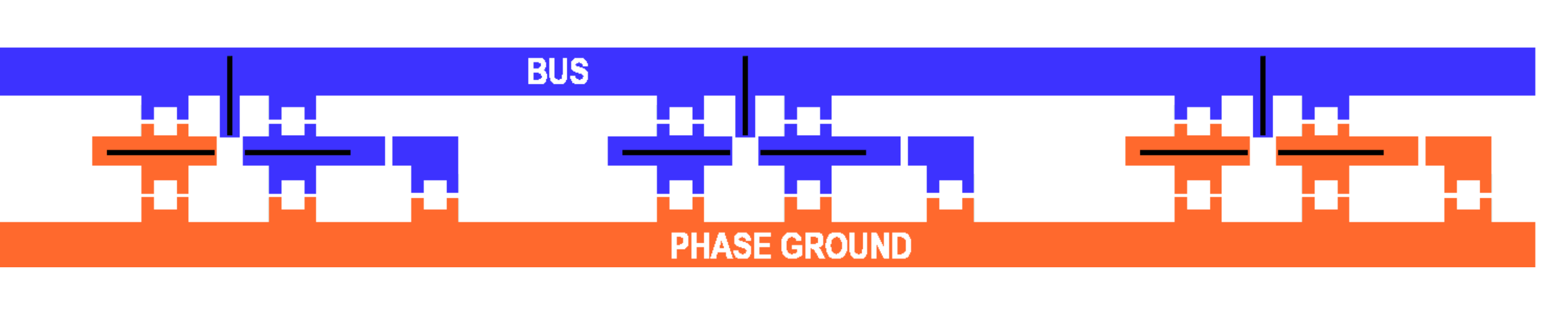}
    \caption{Top four panels: Qubit design showing measurement configurations for $X$, $Y$, $Z$ and Tare ($I$) measurements of a qubit. Bottom panel: portion of a RAMM in a configuration measuring $X\otimes I\otimes Z$.}
    \label{Fig:Qubit}
    \end{figure}

    Once Bell violations have been demonstrated, the next step toward universal quantum computation is a scalable qubit register in which all necessary gate operations (e.g. Clifford gates and the $\pi/8$ phase gate) could be performed. Our phase gate may be easily worked into such a design, as we demonstrate here in a simplification of the random access Majorana memory (RAMM) introduced in Ref.~\onlinecite{Hyart13}:

    As shown in Fig.~\ref{Fig:Qubit}, we can construct a set of islands within the resonator system to function as a single qubit, with measurement settings available to measure any Pauli operator $\{I,X,Y,Z\}$, along with a phase gate that operates around the $Y$ axis. Furthermore, by coupling several qubits to the same register, we may perform \emph{any} Pauli measurement on the qubits. As described in Refs. \onlinecite{Bonderson08b,Bonderson08c}, the set of Clifford gates may be efficiently constructed using Pauli measurements. Combined with the phase gates available on each qubit and the distillation of magic states \cite{Bravyi05} using these phase gates, this design provides the necessary components for universal quantum computation.
    We believe that the method outlined in our Fig.~\ref{Fig:Qubit}, which combines magic state distillation and measurement-only ideas in the context of our proposed phase gate, may very well be the most practical experimental way yet proposed for carrying out universal quantum computation using nanowire MZMs.
\section{Outlook}\label{Sec:Outlook}

    The novel phase gate described in this paper should more than meet the threshold for error correction in carrying out universal fault-tolerant quantum computation using MZMs \cite{Bravyi05}. The real world function of the phase gate, as well as its quantum entanglement properties (beyond the Gottesmann-Knill constraint of pure Clifford operations) can be diagnosed through Bell measurements. These tests of the CHSH inequality are no more daunting than tests of braiding, yet are better targeted toward the eventual implementation of quantum information processing in Majorana-based platforms.  In fact, it has been shown that any operation capable of producing a violation of the CHSH inequality, when combined with Clifford operations, is sufficient for universal quantum computation \cite{Howard12}. By contrast, we do not envision our proposed experiments as tests of quantum nonlocality, as it is unlikely that the qubits in our proposal will be space-like separated. In any case, it is clear that nonlocality is not sufficient for universal quantum computation, as it may be achieved in Ising anyons through braiding alone \cite{Campbell14} (a system that does not even suffice for universal classical computation). The role of the CHSH inequalities as a benchmark in Ising or Majorana systems has been explored before \cite{Zhang07, Brennen09,Howard12,Drummond14}. Here we propose to use this benchmark to experimentally characterize the new phase gate realization we have put forth, a realization that benefits from a relative immunity to timing errors and that can be combined with measurement operations in a unified architecture.

    We believe that our proposal is simple enough, and the consequences of an experimental implementation important enough, that serious consideration should be given by experimentalists toward trying this Bell violation experiment as the very first quantum entanglement experiment in the semiconductor Majorana wire system, even before garden variety braiding experiments are performed. We note here with considerable relief that our phase gate idea leads to a precise test for Bell violation as well as a well-defined platform for universal fault tolerant quantum computation on more or less the same footing, using platforms very similar to what are being currently constructed in laboratories for nanowire MZM braiding experiments.

\section{Acknowledgements}
    We are grateful to Kirill Shtengel for suggesting Bell violations as a diagnostic tool. This work is supported by Microsoft Station Q and LPS-CMTC.
\bibliography{topo-phases}

\begin{thebibliography}{63}
\expandafter\ifx\csname natexlab\endcsname\relax\def\natexlab#1{#1}\fi
\expandafter\ifx\csname bibnamefont\endcsname\relax
  \def\bibnamefont#1{#1}\fi
\expandafter\ifx\csname bibfnamefont\endcsname\relax
  \def\bibfnamefont#1{#1}\fi
\expandafter\ifx\csname citenamefont\endcsname\relax
  \def\citenamefont#1{#1}\fi
\expandafter\ifx\csname url\endcsname\relax
  \def\url#1{\texttt{#1}}\fi
\expandafter\ifx\csname urlprefix\endcsname\relax\def\urlprefix{URL }\fi
\providecommand{\bibinfo}[2]{#2}
\providecommand{\eprint}[2][]{\url{#2}}

\bibitem[{\citenamefont{{Gottesman}}(1998)}]{Gottesman98a}
\bibinfo{author}{\bibfnamefont{D.}~\bibnamefont{{Gottesman}}},
  \bibinfo{journal}{eprint arXiv:quant-ph/9807006}  (\bibinfo{year}{1998}),
  \eprint{quant-ph/9807006}.

\bibitem[{\citenamefont{{Cuffaro}}(2013)}]{Cuffaro13}
\bibinfo{author}{\bibfnamefont{M.~E.} \bibnamefont{{Cuffaro}}},
  \bibinfo{journal}{ArXiv e-prints}  (\bibinfo{year}{2013}),
  \eprint{1310.0938}.

\bibitem[{\citenamefont{Bell}(1964)}]{Bell64}
\bibinfo{author}{\bibfnamefont{J.~S.} \bibnamefont{Bell}},
  \bibinfo{journal}{Physics 1} \textbf{\bibinfo{volume}{3}},
  \bibinfo{pages}{195} (\bibinfo{year}{1964}).

\bibitem[{\citenamefont{Clauser et~al.}(1969)\citenamefont{Clauser, Horne,
  Shimony, and Holt}}]{Clauser69}
\bibinfo{author}{\bibfnamefont{J.~F.} \bibnamefont{Clauser}},
  \bibinfo{author}{\bibfnamefont{M.~A.} \bibnamefont{Horne}},
  \bibinfo{author}{\bibfnamefont{A.}~\bibnamefont{Shimony}}, \bibnamefont{and}
  \bibinfo{author}{\bibfnamefont{R.~A.} \bibnamefont{Holt}},
  \bibinfo{journal}{Phys. Rev. Lett.} \textbf{\bibinfo{volume}{23}},
  \bibinfo{pages}{880} (\bibinfo{year}{1969}),
  \urlprefix\url{http://link.aps.org/doi/10.1103/PhysRevLett.23.880}.

\bibitem[{\citenamefont{Freedman et~al.}(2003)\citenamefont{Freedman, Kitaev,
  Larsen, and Wang}}]{Freedman03a}
\bibinfo{author}{\bibfnamefont{M.}~\bibnamefont{Freedman}},
  \bibinfo{author}{\bibfnamefont{A.}~\bibnamefont{Kitaev}},
  \bibinfo{author}{\bibfnamefont{M.}~\bibnamefont{Larsen}}, \bibnamefont{and}
  \bibinfo{author}{\bibfnamefont{Z.}~\bibnamefont{Wang}},
  \bibinfo{journal}{Bull. Amer. Math. Soc.} \textbf{\bibinfo{volume}{40}},
  \bibinfo{pages}{31} (\bibinfo{year}{2003}).

\bibitem[{\citenamefont{Nayak et~al.}(2008)\citenamefont{Nayak, Simon, Stern,
  Freedman, and Sarma}}]{Nayak08}
\bibinfo{author}{\bibfnamefont{C.}~\bibnamefont{Nayak}},
  \bibinfo{author}{\bibfnamefont{S.~H.} \bibnamefont{Simon}},
  \bibinfo{author}{\bibfnamefont{A.}~\bibnamefont{Stern}},
  \bibinfo{author}{\bibfnamefont{M.}~\bibnamefont{Freedman}}, \bibnamefont{and}
  \bibinfo{author}{\bibfnamefont{S.~D.} \bibnamefont{Sarma}},
  \bibinfo{journal}{Rev. Mod. Phys.} \textbf{\bibinfo{volume}{80}},
  \bibinfo{pages}{1083} (\bibinfo{year}{2008}).

\bibitem[{\citenamefont{Alicea}(2012)}]{Alicea12a}
\bibinfo{author}{\bibfnamefont{J.}~\bibnamefont{Alicea}},
  \bibinfo{journal}{Rep. Prog. Phys.} \textbf{\bibinfo{volume}{75}},
  \bibinfo{pages}{076501} (\bibinfo{year}{2012}), \eprint{arXiv:1202.1293}.

\bibitem[{\citenamefont{{Leijnse} and {Flensberg}}(2012)}]{Leijnse12}
\bibinfo{author}{\bibfnamefont{M.}~\bibnamefont{{Leijnse}}} \bibnamefont{and}
  \bibinfo{author}{\bibfnamefont{K.}~\bibnamefont{{Flensberg}}},
  \bibinfo{journal}{Semiconductor Science Technology}
  \textbf{\bibinfo{volume}{27}}, \bibinfo{eid}{124003} (\bibinfo{year}{2012}),
  \eprint{1206.1736}.

\bibitem[{\citenamefont{Beenakker}(2013)}]{Beenakker2013a}
\bibinfo{author}{\bibfnamefont{C.~W.~J.} \bibnamefont{Beenakker}},
  \bibinfo{journal}{Annu. Rev. Condens. Matter Phys.}
  \textbf{\bibinfo{volume}{4}}, \bibinfo{pages}{113} (\bibinfo{year}{2013}),
  \eprint{arXiv:1112.1950}.

\bibitem[{\citenamefont{{Stanescu} and {Tewari}}(2013)}]{Stanescu13}
\bibinfo{author}{\bibfnamefont{T.~D.} \bibnamefont{{Stanescu}}}
  \bibnamefont{and} \bibinfo{author}{\bibfnamefont{S.}~\bibnamefont{{Tewari}}},
  \bibinfo{journal}{Journal of Physics Condensed Matter}
  \textbf{\bibinfo{volume}{25}}, \bibinfo{eid}{233201} (\bibinfo{year}{2013}),
  \eprint{1302.5433}.

\bibitem[{\citenamefont{{Das Sarma} et~al.}(2015)\citenamefont{{Das Sarma},
  {Freedman}, and {Nayak}}}]{DasSarma15}
\bibinfo{author}{\bibfnamefont{S.}~\bibnamefont{{Das Sarma}}},
  \bibinfo{author}{\bibfnamefont{M.}~\bibnamefont{{Freedman}}},
  \bibnamefont{and} \bibinfo{author}{\bibfnamefont{C.}~\bibnamefont{{Nayak}}},
  \bibinfo{journal}{ArXiv e-prints}  (\bibinfo{year}{2015}),
  \eprint{1501.02813}.

\bibitem[{\citenamefont{{Elliott} and {Franz}}(2015)}]{Elliott15}
\bibinfo{author}{\bibfnamefont{S.~R.} \bibnamefont{{Elliott}}}
  \bibnamefont{and} \bibinfo{author}{\bibfnamefont{M.}~\bibnamefont{{Franz}}},
  \bibinfo{journal}{Reviews of Modern Physics} \textbf{\bibinfo{volume}{87}},
  \bibinfo{pages}{137} (\bibinfo{year}{2015}), \eprint{1403.4976}.

\bibitem[{\citenamefont{Fowler et~al.}(2012)\citenamefont{Fowler, Mariantoni,
  Martinis, and Cleland}}]{Fowler12}
\bibinfo{author}{\bibfnamefont{A.~G.} \bibnamefont{Fowler}},
  \bibinfo{author}{\bibfnamefont{M.}~\bibnamefont{Mariantoni}},
  \bibinfo{author}{\bibfnamefont{J.~M.} \bibnamefont{Martinis}},
  \bibnamefont{and} \bibinfo{author}{\bibfnamefont{A.~N.}
  \bibnamefont{Cleland}}, \bibinfo{journal}{Phys. Rev. A}
  \textbf{\bibinfo{volume}{86}}, \bibinfo{pages}{032324}
  (\bibinfo{year}{2012}),
  \urlprefix\url{http://link.aps.org/doi/10.1103/PhysRevA.86.032324}.

\bibitem[{\citenamefont{Barends et~al.}(2014)\citenamefont{Barends, Kelly,
  Megrant, Veitia, Sank, Jeffrey, White, Mutus, Fowler, Campbell
  et~al.}}]{Martinis14}
\bibinfo{author}{\bibfnamefont{R.}~\bibnamefont{Barends}},
  \bibinfo{author}{\bibfnamefont{J.}~\bibnamefont{Kelly}},
  \bibinfo{author}{\bibfnamefont{A.}~\bibnamefont{Megrant}},
  \bibinfo{author}{\bibfnamefont{A.}~\bibnamefont{Veitia}},
  \bibinfo{author}{\bibfnamefont{D.}~\bibnamefont{Sank}},
  \bibinfo{author}{\bibfnamefont{E.}~\bibnamefont{Jeffrey}},
  \bibinfo{author}{\bibfnamefont{T.~C.} \bibnamefont{White}},
  \bibinfo{author}{\bibfnamefont{J.}~\bibnamefont{Mutus}},
  \bibinfo{author}{\bibfnamefont{A.~G.} \bibnamefont{Fowler}},
  \bibinfo{author}{\bibfnamefont{B.}~\bibnamefont{Campbell}},
  \bibnamefont{et~al.}, \bibinfo{journal}{Nature}
  \textbf{\bibinfo{volume}{508}}, \bibinfo{pages}{500} (\bibinfo{year}{2014}),
  ISSN \bibinfo{issn}{1476-4687},
  \urlprefix\url{http://dx.doi.org/10.1038/nature13171}.

\bibitem[{\citenamefont{{Bonderson} et~al.}(2010)\citenamefont{{Bonderson},
  {Das Sarma}, {Freedman}, and {Nayak}}}]{Bonderson10b}
\bibinfo{author}{\bibfnamefont{P.}~\bibnamefont{{Bonderson}}},
  \bibinfo{author}{\bibfnamefont{S.}~\bibnamefont{{Das Sarma}}},
  \bibinfo{author}{\bibfnamefont{M.}~\bibnamefont{{Freedman}}},
  \bibnamefont{and} \bibinfo{author}{\bibfnamefont{C.}~\bibnamefont{{Nayak}}},
  \bibinfo{journal}{ArXiv e-prints}  (\bibinfo{year}{2010}),
  \eprint{1003.2856}.

\bibitem[{\citenamefont{{Mong} et~al.}(2014)\citenamefont{{Mong}, {Clarke},
  {Alicea}, {Lindner}, {Fendley}, {Nayak}, {Oreg}, {Stern}, {Berg}, {Shtengel}
  et~al.}}]{Mong14a}
\bibinfo{author}{\bibfnamefont{R.~S.~K.} \bibnamefont{{Mong}}},
  \bibinfo{author}{\bibfnamefont{D.~J.} \bibnamefont{{Clarke}}},
  \bibinfo{author}{\bibfnamefont{J.}~\bibnamefont{{Alicea}}},
  \bibinfo{author}{\bibfnamefont{N.~H.} \bibnamefont{{Lindner}}},
  \bibinfo{author}{\bibfnamefont{P.}~\bibnamefont{{Fendley}}},
  \bibinfo{author}{\bibfnamefont{C.}~\bibnamefont{{Nayak}}},
  \bibinfo{author}{\bibfnamefont{Y.}~\bibnamefont{{Oreg}}},
  \bibinfo{author}{\bibfnamefont{A.}~\bibnamefont{{Stern}}},
  \bibinfo{author}{\bibfnamefont{E.}~\bibnamefont{{Berg}}},
  \bibinfo{author}{\bibfnamefont{K.}~\bibnamefont{{Shtengel}}},
  \bibnamefont{et~al.}, \bibinfo{journal}{Physical Review X}
  \textbf{\bibinfo{volume}{4}}, \bibinfo{eid}{011036} (\bibinfo{year}{2014}),
  \eprint{1307.4403}.

\bibitem[{\citenamefont{{Stoudenmire} et~al.}(2015)\citenamefont{{Stoudenmire},
  {Clarke}, {Mong}, and {Alicea}}}]{Stoudenmire15}
\bibinfo{author}{\bibfnamefont{E.~M.} \bibnamefont{{Stoudenmire}}},
  \bibinfo{author}{\bibfnamefont{D.~J.} \bibnamefont{{Clarke}}},
  \bibinfo{author}{\bibfnamefont{R.~S.~K.} \bibnamefont{{Mong}}},
  \bibnamefont{and} \bibinfo{author}{\bibfnamefont{J.}~\bibnamefont{{Alicea}}},
  \bibinfo{journal}{ArXiv e-prints}  (\bibinfo{year}{2015}),
  \eprint{1501.05305}.

\bibitem[{\citenamefont{Barkeshli et~al.}(2013)\citenamefont{Barkeshli, Jian,
  and Qi}}]{Barkeshli13a}
\bibinfo{author}{\bibfnamefont{M.}~\bibnamefont{Barkeshli}},
  \bibinfo{author}{\bibfnamefont{C.-M.} \bibnamefont{Jian}}, \bibnamefont{and}
  \bibinfo{author}{\bibfnamefont{X.-L.} \bibnamefont{Qi}},
  \bibinfo{journal}{Phys. Rev. B} \textbf{\bibinfo{volume}{87}},
  \bibinfo{pages}{045130} (\bibinfo{year}{2013}), \eprint{arXiv:1208.4834}.

\bibitem[{\citenamefont{{Barkeshli} et~al.}()\citenamefont{{Barkeshli},
  {Bonderson}, {Cheng}, and {Wang}}}]{Barkeshli14}
\bibinfo{author}{\bibfnamefont{M.}~\bibnamefont{{Barkeshli}}},
  \bibinfo{author}{\bibfnamefont{P.}~\bibnamefont{{Bonderson}}},
  \bibinfo{author}{\bibfnamefont{M.}~\bibnamefont{{Cheng}}}, \bibnamefont{and}
  \bibinfo{author}{\bibfnamefont{Z.}~\bibnamefont{{Wang}}},
  \emph{\bibinfo{title}{{Symmetry, Defects, and Gauging of Topological
  Phases}}}, \bibinfo{note}{arXiv:1410.4540}.

\bibitem[{\citenamefont{{Kliuchnikov} et~al.}(2014)\citenamefont{{Kliuchnikov},
  {Bocharov}, and {Svore}}}]{Kliuchnikov14}
\bibinfo{author}{\bibfnamefont{V.}~\bibnamefont{{Kliuchnikov}}},
  \bibinfo{author}{\bibfnamefont{A.}~\bibnamefont{{Bocharov}}},
  \bibnamefont{and} \bibinfo{author}{\bibfnamefont{K.~M.}
  \bibnamefont{{Svore}}}, \bibinfo{journal}{Physical Review Letters}
  \textbf{\bibinfo{volume}{112}}, \bibinfo{eid}{140504} (\bibinfo{year}{2014}),
  \eprint{1310.4150}.

\bibitem[{\citenamefont{Terhal}(2015)}]{Terhal13}
\bibinfo{author}{\bibfnamefont{B.~M.} \bibnamefont{Terhal}},
  \bibinfo{journal}{Rev. Mod. Phys.} \textbf{\bibinfo{volume}{87}},
  \bibinfo{pages}{307} (\bibinfo{year}{2015}),
  \urlprefix\url{http://link.aps.org/doi/10.1103/RevModPhys.87.307}.

\bibitem[{\citenamefont{Freedman et~al.}(2006)\citenamefont{Freedman, Nayak,
  and Walker}}]{Freedman06}
\bibinfo{author}{\bibfnamefont{M.}~\bibnamefont{Freedman}},
  \bibinfo{author}{\bibfnamefont{C.}~\bibnamefont{Nayak}}, \bibnamefont{and}
  \bibinfo{author}{\bibfnamefont{K.}~\bibnamefont{Walker}},
  \bibinfo{journal}{Phys. Rev. B} \textbf{\bibinfo{volume}{73}},
  \bibinfo{pages}{245307} (\bibinfo{year}{2006}),
  \urlprefix\url{http://link.aps.org/abstract/PRB/v73/e245307}.

\bibitem[{\citenamefont{{Bravyi} and {Kitaev}}(2005)}]{Bravyi05}
\bibinfo{author}{\bibfnamefont{S.}~\bibnamefont{{Bravyi}}} \bibnamefont{and}
  \bibinfo{author}{\bibfnamefont{A.}~\bibnamefont{{Kitaev}}},
  \bibinfo{journal}{\pra} \textbf{\bibinfo{volume}{71}}, \bibinfo{eid}{022316}
  (\bibinfo{year}{2005}), \eprint{quant-ph/0403025}.

\bibitem[{\citenamefont{{Duclos-Cianci} and {Poulin}}(2015)}]{Duclos-Cianti14}
\bibinfo{author}{\bibfnamefont{G.}~\bibnamefont{{Duclos-Cianci}}}
  \bibnamefont{and} \bibinfo{author}{\bibfnamefont{D.}~\bibnamefont{{Poulin}}},
  \bibinfo{journal}{\pra} \textbf{\bibinfo{volume}{91}}, \bibinfo{eid}{042315}
  (\bibinfo{year}{2015}), \eprint{1403.5280}.

\bibitem[{\citenamefont{Sau et~al.}(2010)\citenamefont{Sau, Tewari, and
  Das~Sarma}}]{Sau10c}
\bibinfo{author}{\bibfnamefont{J.~D.} \bibnamefont{Sau}},
  \bibinfo{author}{\bibfnamefont{S.}~\bibnamefont{Tewari}}, \bibnamefont{and}
  \bibinfo{author}{\bibfnamefont{S.}~\bibnamefont{Das~Sarma}},
  \bibinfo{journal}{Phys. Rev. A} \textbf{\bibinfo{volume}{82}},
  \bibinfo{pages}{052322} (\bibinfo{year}{2010}).

\bibitem[{\citenamefont{{Sau} et~al.}(2011)\citenamefont{{Sau}, {Clarke}, and
  {Tewari}}}]{Sau11b}
\bibinfo{author}{\bibfnamefont{J.~D.} \bibnamefont{{Sau}}},
  \bibinfo{author}{\bibfnamefont{D.~J.} \bibnamefont{{Clarke}}},
  \bibnamefont{and} \bibinfo{author}{\bibfnamefont{S.}~\bibnamefont{{Tewari}}},
  \bibinfo{journal}{\prb} \textbf{\bibinfo{volume}{84}}, \bibinfo{eid}{094505}
  (\bibinfo{year}{2011}), \eprint{1012.0561}.

\bibitem[{\citenamefont{{Alicea} et~al.}(2011)\citenamefont{{Alicea}, {Oreg},
  {Refael}, {von Oppen}, and {Fisher}}}]{Alicea11}
\bibinfo{author}{\bibfnamefont{J.}~\bibnamefont{{Alicea}}},
  \bibinfo{author}{\bibfnamefont{Y.}~\bibnamefont{{Oreg}}},
  \bibinfo{author}{\bibfnamefont{G.}~\bibnamefont{{Refael}}},
  \bibinfo{author}{\bibfnamefont{F.}~\bibnamefont{{von Oppen}}},
  \bibnamefont{and} \bibinfo{author}{\bibfnamefont{M.~P.~A.}
  \bibnamefont{{Fisher}}}, \bibinfo{journal}{Nature Physics}
  \textbf{\bibinfo{volume}{7}}, \bibinfo{pages}{412} (\bibinfo{year}{2011}),
  \eprint{1006.4395}.

\bibitem[{\citenamefont{van Heck et~al.}(2012)\citenamefont{van Heck, Akhmerov,
  Hassler, Burrello, and Beenakker}}]{vanHeck12}
\bibinfo{author}{\bibfnamefont{B.}~\bibnamefont{van Heck}},
  \bibinfo{author}{\bibfnamefont{A.~R.} \bibnamefont{Akhmerov}},
  \bibinfo{author}{\bibfnamefont{F.}~\bibnamefont{Hassler}},
  \bibinfo{author}{\bibfnamefont{M.}~\bibnamefont{Burrello}}, \bibnamefont{and}
  \bibinfo{author}{\bibfnamefont{C.~W.~J.} \bibnamefont{Beenakker}},
  \bibinfo{journal}{New Journal of Physics} \textbf{\bibinfo{volume}{14}},
  \bibinfo{pages}{035019} (\bibinfo{year}{2012}),
  \urlprefix\url{http://stacks.iop.org/1367-2630/14/i=3/a=035019}.

\bibitem[{\citenamefont{{Hyart} et~al.}(2013)\citenamefont{{Hyart}, {van Heck},
  {Fulga}, {Burrello}, {Akhmerov}, and {Beenakker}}}]{Hyart13}
\bibinfo{author}{\bibfnamefont{T.}~\bibnamefont{{Hyart}}},
  \bibinfo{author}{\bibfnamefont{B.}~\bibnamefont{{van Heck}}},
  \bibinfo{author}{\bibfnamefont{I.~C.} \bibnamefont{{Fulga}}},
  \bibinfo{author}{\bibfnamefont{M.}~\bibnamefont{{Burrello}}},
  \bibinfo{author}{\bibfnamefont{A.~R.} \bibnamefont{{Akhmerov}}},
  \bibnamefont{and} \bibinfo{author}{\bibfnamefont{C.~W.~J.}
  \bibnamefont{{Beenakker}}}, \bibinfo{journal}{\prb}
  \textbf{\bibinfo{volume}{88}}, \bibinfo{eid}{035121} (\bibinfo{year}{2013}),
  \eprint{1303.4379}.

\bibitem[{\citenamefont{Kouwenhoven}()}]{Kouwenhoven15}
\bibinfo{author}{\bibfnamefont{L.~P.} \bibnamefont{Kouwenhoven}},
  \bibinfo{howpublished}{private communication}.

\bibitem[{\citenamefont{Marcus}()}]{Marcus15}
\bibinfo{author}{\bibfnamefont{C.~M.} \bibnamefont{Marcus}},
  \bibinfo{howpublished}{private communication}.

\bibitem[{\citenamefont{Clarke et~al.}(2013)\citenamefont{Clarke, Alicea, and
  Shtengel}}]{Clarke13a}
\bibinfo{author}{\bibfnamefont{D.~J.} \bibnamefont{Clarke}},
  \bibinfo{author}{\bibfnamefont{J.}~\bibnamefont{Alicea}}, \bibnamefont{and}
  \bibinfo{author}{\bibfnamefont{K.}~\bibnamefont{Shtengel}},
  \bibinfo{journal}{Nat. Commun.} \textbf{\bibinfo{volume}{4}},
  \bibinfo{pages}{1348} (\bibinfo{year}{2013}), \eprint{arXiv:1204.5479}.

\bibitem[{\citenamefont{Lindner et~al.}(2012)\citenamefont{Lindner, Berg,
  Refael, and Stern}}]{Lindner12}
\bibinfo{author}{\bibfnamefont{N.~H.} \bibnamefont{Lindner}},
  \bibinfo{author}{\bibfnamefont{E.}~\bibnamefont{Berg}},
  \bibinfo{author}{\bibfnamefont{G.}~\bibnamefont{Refael}}, \bibnamefont{and}
  \bibinfo{author}{\bibfnamefont{A.}~\bibnamefont{Stern}},
  \bibinfo{journal}{Phys. Rev. X} \textbf{\bibinfo{volume}{2}},
  \bibinfo{pages}{041002} (\bibinfo{year}{2012}), \eprint{arXiv:1204.5733}.

\bibitem[{\citenamefont{Cheng}(2012)}]{Cheng12}
\bibinfo{author}{\bibfnamefont{M.}~\bibnamefont{Cheng}},
  \bibinfo{journal}{Phys. Rev. B} \textbf{\bibinfo{volume}{86}},
  \bibinfo{pages}{195126} (\bibinfo{year}{2012}), \eprint{arXiv:1204.6084}.

\bibitem[{\citenamefont{Bonderson et~al.}(2010)\citenamefont{Bonderson, Clarke,
  Nayak, and Shtengel}}]{Bonderson10a}
\bibinfo{author}{\bibfnamefont{P.}~\bibnamefont{Bonderson}},
  \bibinfo{author}{\bibfnamefont{D.~J.} \bibnamefont{Clarke}},
  \bibinfo{author}{\bibfnamefont{C.}~\bibnamefont{Nayak}}, \bibnamefont{and}
  \bibinfo{author}{\bibfnamefont{K.}~\bibnamefont{Shtengel}},
  \bibinfo{journal}{Phys. Rev. Lett.} \textbf{\bibinfo{volume}{104}},
  \bibinfo{pages}{180505} (\bibinfo{year}{2010}), \eprint{arXiv:0911.2691}.

\bibitem[{\citenamefont{{Clarke} and {Shtengel}}(2010)}]{Clarke10}
\bibinfo{author}{\bibfnamefont{D.~J.} \bibnamefont{{Clarke}}} \bibnamefont{and}
  \bibinfo{author}{\bibfnamefont{K.}~\bibnamefont{{Shtengel}}},
  \bibinfo{journal}{\prb} \textbf{\bibinfo{volume}{82}}, \bibinfo{eid}{180519}
  (\bibinfo{year}{2010}), \eprint{1009.0302}.

\bibitem[{\citenamefont{{Hassler} et~al.}(2010)\citenamefont{{Hassler},
  {Akhmerov}, {Hou}, and {Beenakker}}}]{Hassler10}
\bibinfo{author}{\bibfnamefont{F.}~\bibnamefont{{Hassler}}},
  \bibinfo{author}{\bibfnamefont{A.~R.} \bibnamefont{{Akhmerov}}},
  \bibinfo{author}{\bibfnamefont{C.-Y.} \bibnamefont{{Hou}}}, \bibnamefont{and}
  \bibinfo{author}{\bibfnamefont{C.~W.~J.} \bibnamefont{{Beenakker}}},
  \bibinfo{journal}{New Journal of Physics} \textbf{\bibinfo{volume}{12}},
  \bibinfo{pages}{125002} (\bibinfo{year}{2010}), \eprint{1005.3423}.

\bibitem[{\citenamefont{{Hassler} et~al.}(2011)\citenamefont{{Hassler},
  {Akhmerov}, and {Beenakker}}}]{Hassler11}
\bibinfo{author}{\bibfnamefont{F.}~\bibnamefont{{Hassler}}},
  \bibinfo{author}{\bibfnamefont{A.~R.} \bibnamefont{{Akhmerov}}},
  \bibnamefont{and} \bibinfo{author}{\bibfnamefont{C.~W.~J.}
  \bibnamefont{{Beenakker}}}, \bibinfo{journal}{New Journal of Physics}
  \textbf{\bibinfo{volume}{13}}, \bibinfo{eid}{095004} (\bibinfo{year}{2011}),
  \eprint{1105.0315}.

\bibitem[{\citenamefont{{Bonderson} and {Lutchyn}}(2011)}]{Bonderson11b}
\bibinfo{author}{\bibfnamefont{P.}~\bibnamefont{{Bonderson}}} \bibnamefont{and}
  \bibinfo{author}{\bibfnamefont{R.~M.} \bibnamefont{{Lutchyn}}},
  \bibinfo{journal}{Phys. Rev. Lett.} \textbf{\bibinfo{volume}{106}},
  \bibinfo{pages}{130505} (\bibinfo{year}{2011}), \eprint{1011.1784}.

\bibitem[{\citenamefont{{Bonderson} et~al.}(2013)\citenamefont{{Bonderson},
  {Fidkowski}, {Freedman}, and {Walker}}}]{Bonderson13b}
\bibinfo{author}{\bibfnamefont{P.}~\bibnamefont{{Bonderson}}},
  \bibinfo{author}{\bibfnamefont{L.}~\bibnamefont{{Fidkowski}}},
  \bibinfo{author}{\bibfnamefont{M.}~\bibnamefont{{Freedman}}},
  \bibnamefont{and} \bibinfo{author}{\bibfnamefont{K.}~\bibnamefont{{Walker}}},
  \bibinfo{journal}{ArXiv e-prints}  (\bibinfo{year}{2013}),
  \eprint{1306.2379}.

\bibitem[{\citenamefont{Mourik et~al.}(2012)\citenamefont{Mourik, Zuo, Frolov,
  Plissard, Bakkers, and Kouwenhoven}}]{Mourik12}
\bibinfo{author}{\bibfnamefont{V.}~\bibnamefont{Mourik}},
  \bibinfo{author}{\bibfnamefont{K.}~\bibnamefont{Zuo}},
  \bibinfo{author}{\bibfnamefont{S.~M.} \bibnamefont{Frolov}},
  \bibinfo{author}{\bibfnamefont{S.~R.} \bibnamefont{Plissard}},
  \bibinfo{author}{\bibfnamefont{E.~P. A.~M.} \bibnamefont{Bakkers}},
  \bibnamefont{and} \bibinfo{author}{\bibfnamefont{L.~P.}
  \bibnamefont{Kouwenhoven}}, \bibinfo{journal}{Science}
  \textbf{\bibinfo{volume}{336}}, \bibinfo{pages}{1003} (\bibinfo{year}{2012}).

\bibitem[{\citenamefont{Das et~al.}(2012)\citenamefont{Das, Ronen, Most, Oreg,
  Heiblum, and Shtrikman}}]{Das12}
\bibinfo{author}{\bibfnamefont{A.}~\bibnamefont{Das}},
  \bibinfo{author}{\bibfnamefont{Y.}~\bibnamefont{Ronen}},
  \bibinfo{author}{\bibfnamefont{Y.}~\bibnamefont{Most}},
  \bibinfo{author}{\bibfnamefont{Y.}~\bibnamefont{Oreg}},
  \bibinfo{author}{\bibfnamefont{M.}~\bibnamefont{Heiblum}}, \bibnamefont{and}
  \bibinfo{author}{\bibfnamefont{H.}~\bibnamefont{Shtrikman}},
  \bibinfo{journal}{Nat. Phys.} \textbf{\bibinfo{volume}{8}},
  \bibinfo{pages}{887} (\bibinfo{year}{2012}), \eprint{arXiv:1205.7073}.

\bibitem[{\citenamefont{Deng et~al.}(2012)\citenamefont{Deng, Yu, Huang,
  Larsson, Caroff, and Xu}}]{Deng12}
\bibinfo{author}{\bibfnamefont{M.~T.} \bibnamefont{Deng}},
  \bibinfo{author}{\bibfnamefont{C.~L.} \bibnamefont{Yu}},
  \bibinfo{author}{\bibfnamefont{G.~Y.} \bibnamefont{Huang}},
  \bibinfo{author}{\bibfnamefont{M.}~\bibnamefont{Larsson}},
  \bibinfo{author}{\bibfnamefont{P.}~\bibnamefont{Caroff}}, \bibnamefont{and}
  \bibinfo{author}{\bibfnamefont{H.~Q.} \bibnamefont{Xu}},
  \bibinfo{journal}{Nano Letters} \textbf{\bibinfo{volume}{12}},
  \bibinfo{pages}{6414} (\bibinfo{year}{2012}), \eprint{arXiv:1204.4130}.

\bibitem[{\citenamefont{Rokhinson et~al.}(2012)\citenamefont{Rokhinson, Liu,
  and Furdyna}}]{Rokhinson12}
\bibinfo{author}{\bibfnamefont{L.~P.} \bibnamefont{Rokhinson}},
  \bibinfo{author}{\bibfnamefont{X.}~\bibnamefont{Liu}}, \bibnamefont{and}
  \bibinfo{author}{\bibfnamefont{J.~K.} \bibnamefont{Furdyna}},
  \bibinfo{journal}{Nat. Phys.} \textbf{\bibinfo{volume}{8}},
  \bibinfo{pages}{795} (\bibinfo{year}{2012}), ISSN \bibinfo{issn}{1745-2473},
  \eprint{arXiv:1204.4212}.

\bibitem[{\citenamefont{Finck et~al.}(2013)\citenamefont{Finck, Van~Harlingen,
  Mohseni, Jung, and Li}}]{Finck12}
\bibinfo{author}{\bibfnamefont{A.~D.~K.} \bibnamefont{Finck}},
  \bibinfo{author}{\bibfnamefont{D.~J.} \bibnamefont{Van~Harlingen}},
  \bibinfo{author}{\bibfnamefont{P.~K.} \bibnamefont{Mohseni}},
  \bibinfo{author}{\bibfnamefont{K.}~\bibnamefont{Jung}}, \bibnamefont{and}
  \bibinfo{author}{\bibfnamefont{X.}~\bibnamefont{Li}}, \bibinfo{journal}{Phys.
  Rev. Lett.} \textbf{\bibinfo{volume}{110}}, \bibinfo{pages}{126406}
  (\bibinfo{year}{2013}).

\bibitem[{\citenamefont{Churchill et~al.}(2013)\citenamefont{Churchill, Fatemi,
  Grove-Rasmussen, Deng, Caroff, Xu, and Marcus}}]{Churchill13}
\bibinfo{author}{\bibfnamefont{H.~O.~H.} \bibnamefont{Churchill}},
  \bibinfo{author}{\bibfnamefont{V.}~\bibnamefont{Fatemi}},
  \bibinfo{author}{\bibfnamefont{K.}~\bibnamefont{Grove-Rasmussen}},
  \bibinfo{author}{\bibfnamefont{M.~T.} \bibnamefont{Deng}},
  \bibinfo{author}{\bibfnamefont{P.}~\bibnamefont{Caroff}},
  \bibinfo{author}{\bibfnamefont{H.~Q.} \bibnamefont{Xu}}, \bibnamefont{and}
  \bibinfo{author}{\bibfnamefont{C.~M.} \bibnamefont{Marcus}},
  \bibinfo{journal}{Phys. Rev. B} \textbf{\bibinfo{volume}{87}},
  \bibinfo{pages}{241401} (\bibinfo{year}{2013}).

\bibitem[{\citenamefont{{Chang} et~al.}(2014)\citenamefont{{Chang}, {Albrecht},
  {Jespersen}, {Kuemmeth}, {Krogstrup}, {Nyg{\aa}rd}, and {Marcus}}}]{Chang14}
\bibinfo{author}{\bibfnamefont{W.}~\bibnamefont{{Chang}}},
  \bibinfo{author}{\bibfnamefont{S.~M.} \bibnamefont{{Albrecht}}},
  \bibinfo{author}{\bibfnamefont{T.~S.} \bibnamefont{{Jespersen}}},
  \bibinfo{author}{\bibfnamefont{F.}~\bibnamefont{{Kuemmeth}}},
  \bibinfo{author}{\bibfnamefont{P.}~\bibnamefont{{Krogstrup}}},
  \bibinfo{author}{\bibfnamefont{J.}~\bibnamefont{{Nyg{\aa}rd}}},
  \bibnamefont{and} \bibinfo{author}{\bibfnamefont{C.~M.}
  \bibnamefont{{Marcus}}}, \bibinfo{journal}{ArXiv e-prints}
  (\bibinfo{year}{2014}), \eprint{1411.6255}.

\bibitem[{\citenamefont{{Lutchyn} et~al.}(2010)\citenamefont{{Lutchyn}, {Sau},
  and {Das Sarma}}}]{Lutchyn10}
\bibinfo{author}{\bibfnamefont{R.~M.} \bibnamefont{{Lutchyn}}},
  \bibinfo{author}{\bibfnamefont{J.~D.} \bibnamefont{{Sau}}}, \bibnamefont{and}
  \bibinfo{author}{\bibfnamefont{S.}~\bibnamefont{{Das Sarma}}},
  \bibinfo{journal}{Phys. Rev. Lett.} \textbf{\bibinfo{volume}{105}},
  \bibinfo{pages}{077001} (\bibinfo{year}{2010}), \eprint{1002.4033}.

\bibitem[{\citenamefont{{Oreg} et~al.}(2010)\citenamefont{{Oreg}, {Refael}, and
  {von Oppen}}}]{Oreg10}
\bibinfo{author}{\bibfnamefont{Y.}~\bibnamefont{{Oreg}}},
  \bibinfo{author}{\bibfnamefont{G.}~\bibnamefont{{Refael}}}, \bibnamefont{and}
  \bibinfo{author}{\bibfnamefont{F.}~\bibnamefont{{von Oppen}}},
  \bibinfo{journal}{Phys. Rev. Lett.} \textbf{\bibinfo{volume}{105}},
  \bibinfo{pages}{177002} (\bibinfo{year}{2010}), \eprint{1003.1145}.

\bibitem[{\citenamefont{{Sau} et~al.}(2010)\citenamefont{{Sau}, {Tewari},
  {Lutchyn}, {Stanescu}, and {Das Sarma}}}]{Sau10b}
\bibinfo{author}{\bibfnamefont{J.~D.} \bibnamefont{{Sau}}},
  \bibinfo{author}{\bibfnamefont{S.}~\bibnamefont{{Tewari}}},
  \bibinfo{author}{\bibfnamefont{R.~M.} \bibnamefont{{Lutchyn}}},
  \bibinfo{author}{\bibfnamefont{T.~D.} \bibnamefont{{Stanescu}}},
  \bibnamefont{and} \bibinfo{author}{\bibfnamefont{S.}~\bibnamefont{{Das
  Sarma}}}, \bibinfo{journal}{\prb} \textbf{\bibinfo{volume}{82}},
  \bibinfo{eid}{214509} (\bibinfo{year}{2010}), \eprint{1006.2829}.

\bibitem[{\citenamefont{Aharonov and Casher}(1984)}]{Aharanov84}
\bibinfo{author}{\bibfnamefont{Y.}~\bibnamefont{Aharonov}} \bibnamefont{and}
  \bibinfo{author}{\bibfnamefont{A.}~\bibnamefont{Casher}},
  \bibinfo{journal}{Phys. Rev. Lett.} \textbf{\bibinfo{volume}{53}},
  \bibinfo{pages}{319} (\bibinfo{year}{1984}),
  \urlprefix\url{http://link.aps.org/doi/10.1103/PhysRevLett.53.319}.

\bibitem[{\citenamefont{Elion et~al.}(1993)\citenamefont{Elion, Wachters, Sohn,
  and Mooij}}]{Elion93}
\bibinfo{author}{\bibfnamefont{W.~J.} \bibnamefont{Elion}},
  \bibinfo{author}{\bibfnamefont{J.~J.} \bibnamefont{Wachters}},
  \bibinfo{author}{\bibfnamefont{L.~L.} \bibnamefont{Sohn}}, \bibnamefont{and}
  \bibinfo{author}{\bibfnamefont{J.~E.} \bibnamefont{Mooij}},
  \bibinfo{journal}{Phys. Rev. Lett.} \textbf{\bibinfo{volume}{71}},
  \bibinfo{pages}{2311} (\bibinfo{year}{1993}),
  \urlprefix\url{http://link.aps.org/doi/10.1103/PhysRevLett.71.2311}.

\bibitem[{\citenamefont{K\"onig et~al.}(2006)\citenamefont{K\"onig,
  Tschetschetkin, Hankiewicz, Sinova, Hock, Daumer, Sch\"afer, Becker, Buhmann,
  and Molenkamp}}]{Konig06}
\bibinfo{author}{\bibfnamefont{M.}~\bibnamefont{K\"onig}},
  \bibinfo{author}{\bibfnamefont{A.}~\bibnamefont{Tschetschetkin}},
  \bibinfo{author}{\bibfnamefont{E.~M.} \bibnamefont{Hankiewicz}},
  \bibinfo{author}{\bibfnamefont{J.}~\bibnamefont{Sinova}},
  \bibinfo{author}{\bibfnamefont{V.}~\bibnamefont{Hock}},
  \bibinfo{author}{\bibfnamefont{V.}~\bibnamefont{Daumer}},
  \bibinfo{author}{\bibfnamefont{M.}~\bibnamefont{Sch\"afer}},
  \bibinfo{author}{\bibfnamefont{C.~R.} \bibnamefont{Becker}},
  \bibinfo{author}{\bibfnamefont{H.}~\bibnamefont{Buhmann}}, \bibnamefont{and}
  \bibinfo{author}{\bibfnamefont{L.~W.} \bibnamefont{Molenkamp}},
  \bibinfo{journal}{Phys. Rev. Lett.} \textbf{\bibinfo{volume}{96}},
  \bibinfo{pages}{076804} (\bibinfo{year}{2006}),
  \urlprefix\url{http://link.aps.org/doi/10.1103/PhysRevLett.96.076804}.

\bibitem[{\citenamefont{Grosfeld and Stern}(2011)}]{Grosfeld11}
\bibinfo{author}{\bibfnamefont{E.}~\bibnamefont{Grosfeld}} \bibnamefont{and}
  \bibinfo{author}{\bibfnamefont{A.}~\bibnamefont{Stern}},
  \bibinfo{journal}{Proceedings of the National Academy of Sciences}
  \textbf{\bibinfo{volume}{108}}, \bibinfo{pages}{11810}
  (\bibinfo{year}{2011}),
  \eprint{http://www.pnas.org/content/108/29/11810.full.pdf},
  \urlprefix\url{http://www.pnas.org/content/108/29/11810.abstract}.

\bibitem[{\citenamefont{Resta}(1994)}]{Resta94}
\bibinfo{author}{\bibfnamefont{R.}~\bibnamefont{Resta}}, \bibinfo{journal}{Rev.
  Mod. Phys.} \textbf{\bibinfo{volume}{66}}, \bibinfo{pages}{899}
  (\bibinfo{year}{1994}),
  \urlprefix\url{http://link.aps.org/doi/10.1103/RevModPhys.66.899}.

\bibitem[{\citenamefont{Koch et~al.}(2007)\citenamefont{Koch, Yu, Gambetta,
  Houck, Schuster, Majer, Blais, Devoret, Girvin, and Schoelkopf}}]{Koch07}
\bibinfo{author}{\bibfnamefont{J.}~\bibnamefont{Koch}},
  \bibinfo{author}{\bibfnamefont{T.~M.} \bibnamefont{Yu}},
  \bibinfo{author}{\bibfnamefont{J.}~\bibnamefont{Gambetta}},
  \bibinfo{author}{\bibfnamefont{A.~A.} \bibnamefont{Houck}},
  \bibinfo{author}{\bibfnamefont{D.~I.} \bibnamefont{Schuster}},
  \bibinfo{author}{\bibfnamefont{J.}~\bibnamefont{Majer}},
  \bibinfo{author}{\bibfnamefont{A.}~\bibnamefont{Blais}},
  \bibinfo{author}{\bibfnamefont{M.~H.} \bibnamefont{Devoret}},
  \bibinfo{author}{\bibfnamefont{S.~M.} \bibnamefont{Girvin}},
  \bibnamefont{and} \bibinfo{author}{\bibfnamefont{R.~J.}
  \bibnamefont{Schoelkopf}}, \bibinfo{journal}{Phys. Rev. A}
  \textbf{\bibinfo{volume}{76}}, \bibinfo{pages}{042319}
  (\bibinfo{year}{2007}),
  \urlprefix\url{http://link.aps.org/doi/10.1103/PhysRevA.76.042319}.

\bibitem[{\citenamefont{Bonderson et~al.}(2008)\citenamefont{Bonderson,
  Freedman, and Nayak}}]{Bonderson08b}
\bibinfo{author}{\bibfnamefont{P.}~\bibnamefont{Bonderson}},
  \bibinfo{author}{\bibfnamefont{M.}~\bibnamefont{Freedman}}, \bibnamefont{and}
  \bibinfo{author}{\bibfnamefont{C.}~\bibnamefont{Nayak}},
  \bibinfo{journal}{Phys. Rev. Lett.} \textbf{\bibinfo{volume}{101}},
  \bibinfo{pages}{010501} (\bibinfo{year}{2008}), \eprint{arXiv:0802.0279}.

\bibitem[{\citenamefont{Bonderson et~al.}(2009)\citenamefont{Bonderson,
  Freedman, and Nayak}}]{Bonderson08c}
\bibinfo{author}{\bibfnamefont{P.}~\bibnamefont{Bonderson}},
  \bibinfo{author}{\bibfnamefont{M.}~\bibnamefont{Freedman}}, \bibnamefont{and}
  \bibinfo{author}{\bibfnamefont{C.}~\bibnamefont{Nayak}},
  \bibinfo{journal}{Annals of Physics} \textbf{\bibinfo{volume}{324}},
  \bibinfo{pages}{787} (\bibinfo{year}{2009}), \eprint{arXiv:0808.1933}.

\bibitem[{\citenamefont{{Howard} and {Vala}}(2012)}]{Howard12}
\bibinfo{author}{\bibfnamefont{M.}~\bibnamefont{{Howard}}} \bibnamefont{and}
  \bibinfo{author}{\bibfnamefont{J.}~\bibnamefont{{Vala}}},
  \bibinfo{journal}{\pra} \textbf{\bibinfo{volume}{85}}, \bibinfo{eid}{022304}
  (\bibinfo{year}{2012}), \eprint{1112.1516}.

\bibitem[{\citenamefont{{Campbell} et~al.}(2014)\citenamefont{{Campbell},
  {Hoban}, and {Eisert}}}]{Campbell14}
\bibinfo{author}{\bibfnamefont{E.~T.} \bibnamefont{{Campbell}}},
  \bibinfo{author}{\bibfnamefont{M.~J.} \bibnamefont{{Hoban}}},
  \bibnamefont{and} \bibinfo{author}{\bibfnamefont{J.}~\bibnamefont{{Eisert}}},
  \bibinfo{journal}{Quant. Inf. Comp.} \textbf{\bibinfo{volume}{14}},
  \bibinfo{pages}{0981} (\bibinfo{year}{2014}).

\bibitem[{\citenamefont{Zhang et~al.}(2007)\citenamefont{Zhang, Tewari, and
  Das~Sarma}}]{Zhang07}
\bibinfo{author}{\bibfnamefont{C.}~\bibnamefont{Zhang}},
  \bibinfo{author}{\bibfnamefont{S.}~\bibnamefont{Tewari}}, \bibnamefont{and}
  \bibinfo{author}{\bibfnamefont{S.}~\bibnamefont{Das~Sarma}},
  \bibinfo{journal}{Phys. Rev. Lett.} \textbf{\bibinfo{volume}{99}},
  \bibinfo{pages}{220502} (\bibinfo{year}{2007}),
  \urlprefix\url{http://link.aps.org/doi/10.1103/PhysRevLett.99.220502}.

\bibitem[{\citenamefont{Brennen et~al.}(2009)\citenamefont{Brennen, Iblisdir,
  Pachos, and Slingerland}}]{Brennen09}
\bibinfo{author}{\bibfnamefont{G.~K.} \bibnamefont{Brennen}},
  \bibinfo{author}{\bibfnamefont{S.}~\bibnamefont{Iblisdir}},
  \bibinfo{author}{\bibfnamefont{J.~K.} \bibnamefont{Pachos}},
  \bibnamefont{and} \bibinfo{author}{\bibfnamefont{J.~K.}
  \bibnamefont{Slingerland}}, \bibinfo{journal}{New Journal of Physics}
  \textbf{\bibinfo{volume}{11}}, \bibinfo{pages}{103023}
  (\bibinfo{year}{2009}),
  \urlprefix\url{http://stacks.iop.org/1367-2630/11/i=10/a=103023}.

\bibitem[{\citenamefont{Drummond et~al.}(2014)\citenamefont{Drummond, Kovalev,
  Hou, Shtengel, and Pryadko}}]{Drummond14}
\bibinfo{author}{\bibfnamefont{D.~E.} \bibnamefont{Drummond}},
  \bibinfo{author}{\bibfnamefont{A.~A.} \bibnamefont{Kovalev}},
  \bibinfo{author}{\bibfnamefont{C.-Y.} \bibnamefont{Hou}},
  \bibinfo{author}{\bibfnamefont{K.}~\bibnamefont{Shtengel}}, \bibnamefont{and}
  \bibinfo{author}{\bibfnamefont{L.~P.} \bibnamefont{Pryadko}},
  \bibinfo{journal}{Phys. Rev. B} \textbf{\bibinfo{volume}{90}},
  \bibinfo{pages}{115404} (\bibinfo{year}{2014}),
  \urlprefix\url{http://link.aps.org/doi/10.1103/PhysRevB.90.115404}.

\end{thebibliography}
\end{document}